\documentclass[aps,twocolumn,floatfix,titlepage,balancelastpage,raggedbottom,amsfonts,amssymb,amsmath,showpacs]{revtex4}
\usepackage{epsfig}
\usepackage{bm}
\usepackage{times}
\usepackage{mathptmx}
\usepackage{amsmath}
\usepackage{graphicx}
\usepackage{graphics}
\usepackage{amssymb}
\usepackage{bm}
\usepackage{multirow}
\usepackage[english]{babel}
\begin{document}
%=======================================================================
\title{Coarse-graining MARTINI model for molecular-dynamics simulations of 
the wetting properties of graphitic surfaces with non-ionic, long-chain and 
T-shaped surfactants}
\author{Danilo Sergi}
\affiliation{University of Applied Sciences (SUPSI), 
The iCIMSI Research Institute, 
Galleria 2, CH-6928 Manno, Switzerland}
\author{Giulio Scocchi}
\affiliation{University of Applied Sciences (SUPSI), 
The iCIMSI Research Institute, 
Galleria 2, CH-6928 Manno, Switzerland}
\author{Alberto Ortona}
\affiliation{University of Applied Sciences (SUPSI), 
The iCIMSI Research Institute, 
Galleria 2, CH-6928 Manno, Switzerland}

\date{\today}

%% 47.11.Mn Molecular dynamics methods  
%% 47.55.dk Surfactant effects  
%% 47.55.dr Interactions with surfaces  
%% 47.55.np Contact lines  
%================================================
\begin{abstract}
We report on a molecular dynamics investigation of the wetting properties of graphitic surfaces 
by various solutions at concentrations $1-8$ wt$\%$ of commercially available non-ionic surfactants 
with long hydrophilic chains, linear or T-shaped. These are surfactants of length up to $160$ [\AA].
It turns out that molecular dynamics simulations of such systems ask 
for a number of solvent particles that can be reached without seriously compromising
computational efficiency only by employing a coarse-grained model. The MARTINI force field 
with polarizable water offers a framework particularly suited for the parameterization of our systems. 
In general, its advantages over other coarse-grained models are the possibility to explore faster 
long time scales and the wider range of applicability. Although the accuracy is sometimes put under question,
the results for the wetting properties by pure water are in good agreement with those for 
the corresponding atomistic systems and theoretical predictions. On the other hand, the bulk properties 
of various aqueous surfactant solutions indicate that the micellar formation process is too strong.
For this reason, a typical experimental configuration is better approached by preparing 
the droplets with the surfactants arranged in the initial state in the vicinity of contact line.
Cross-comparisons are possible and illuminating, but equilibrium contanct angles as obtained 
from simulations overestimate the experimental results. Nevertheless, our findings can provide guidelines 
for the preliminary assessment and screening of surfactants. Most importantly, it is found that the wetting
properties mainly depend on the length and apolarity of the hydrophobic tail, for linear surfactants, and the length
of the hydrophilic headgroup for T-shaped surfactants. Moreover, the T-shaped topology 
appears to favor the adsorption of surfactants onto the graphitic surface and faster spreading.
\end{abstract}
%=======================================================================
\maketitle
%======================================================================

\section{Introduction}

Molecular dynamics (MD) simulations have become an important complement to 
experimental and theoretical work across a variety of disciplines. Often,
one of the encountered obstacles to further advances are the prohibitive computational
resources necessary for a description at the molecular level of systems
containing a large number of particles \cite{shinoda_cg}. This situation is in general quite
common when water is involved as solvent. One way to overcome this limitation is to reduce
the degrees of freedom to the essential for the issues under investigation. This
procedure is referred to as coarse-graining because the system is suitably organized
into units identifying groups of atoms. To achieve this, two schemes are
basically available \cite{shinoda1,shinoda2,martini1,martini2}. 
The differences mainly reside in the decomposition of the system into building 
blocks and the selected properties 
according to which inter-particle interactions are calibrated.
Our work is based on the MARTINI force field because allowing 
to build readily and systematically systems of coarse-grained (CG) units 
in a prescribed way \cite{martini1}. Within this framework four heavy atoms are 
generally mapped into one interaction site. 
The resulting beads are classified into four main categories that separate further 
into subtypes. Lennard-Jones (LJ) interactions are organized into ten levels of
strength and evaluated with a function of type $12$-$6$. In its original formulation, 
the MARTINI model does not contemplate solid phases \cite{martini1}. Following 
empirical arguments, we treat graphitic surfaces and their wetting properties 
by surfactant solutions. In that respect, the surface tension of water is
of course a key property. The polarizable MARTINI model for water \cite{martini2} 
still underestimates the experimental value. The expected balance of forces at the
interface with the graphitic surface is approximately restored by tuning the interaction
strength of the relative LJ potential. In this way, we first
reproduce in the macroscopic limit the contact angle for pure water from measurements 
at present carried out under the most ideal conditions \cite{wang}. Interestingly, 
our results correlate well with those using atomistic models for water consistent with
the experimental surface tension \cite{epsilon,contact}. This allows us to conclude that the
MARTINI force field can provide an effective framework for investigating
the wetting properties of surfaces. Surfactants are molecules exhibiting
opposite behavior toward water at their extremes, namely hydrophilic in the 
headgroup and hydrophobic in the tail group. Importantly, these competing 
attributes are responsible for the decrease of the surface tension of 
the resulting solution \cite{rosen}. In practice, this means that an aqueous 
solution would wet better a surface with the addition of surfactants. At higher concentrations,
another important property related to the amphiphilic nature of surfactants is their 
ability to self-assemble into micelles \cite{rosen}. Applications in the industry are innumerable,
including printing, detergency, oil recovery, dispersion, emulsification, disinfection, etc.~\cite{rosen}.
In this work, we concentrate on solutions with non-ionic surfactants. As conveyed by the title, 
emphasis is put on the wetting properties by measuring the contact angle of droplets.
The identity of the surfactants under investigation is confidential for industrial reasons. 
Specifically, their use as wetting agents is necessary for the optimal dispersion of reinforcement 
fillers in polymeric matrices (low chemical compatibility to each other) \cite{dispersion}. 
This phase is rather promiment in the manufacturing route of composite materials in order to meet challenging, 
target properties \cite{canada,delaware}.
Our work aims at providing useful guidance for planning applications.
So, the basic mechanisms governing wetting phenomena by aqueous surfactant solutions are at the
center of our attention. The strategy consists in simulating six representative surfactants
differing in simple attributes such as the length of the hydrophilic and hydrophobic groups
or the topology (linear and T-shaped). Our research develops as follows. In the next
Section, the MARTINI representation of all the constituents of the systems is introduced.
Section \ref{sec:sim} is concerned with the general MD simulation settings. In Sec.~\ref{sec:results}
are presented the results for systems of pure water and with surfactants: first the bulk
properties of solutions and then the wetting behavior on graphitic surfaces.
The last Section is devoted to conclusive remarks.

%======================================================================

\section{Coarse-graining of the systems}\label{sec:cg}

\textit{Force field}.~The MARTINI force field is designed in order to account
for a variety of chemical compounds preventing the need to recalibrate the
various interaction parameters for every application \cite{martini1,martini2}. 
The idea is to classify atomic functional groups into four main categories: 
polar, nonpolar, apolar 
and charged. Every category subdivides further into subtypes so as to represent
more finely the underlying structural properties. Ten levels of interaction
are possible between the diverse types of CG particles. More precisely, inter-particle
LJ forces are derived from the potential energy
\begin{equation*}
U_{\mathrm{LJ}}(r_{ij})=4\varepsilon_{ij}\Big[\Big(\frac{\sigma_{ij}}{r_{ij}}\Big)^{12}
-\Big(\frac{\sigma_{ij}}{r_{ij}}\Big)^{6}\Big]\ .
\end{equation*}
The indices $i$ and $j$ label the particles. $r_{ij}$ is the distance separating pairs
of particles. The parameter $\varepsilon_{ij}$ adjusts the strength of the forces
according to the available levels of interaction. The coefficient $\sigma_{ij}$ is
related to the equilibrium distance between particles via the formula $2^{1/6}\sigma_{ij}$.
If $r_{ij}<2^{1/6}\sigma_{ij}$ the force is repulsive, otherwise it is attractive.
The parameter $\sigma_{ij}$ is also usually referred to as molecular diameter.
Electrostatic interactions are computed from the coulombic potential
\begin{equation*}
U_{\mathrm{C}}(r_{ij})=\frac{q_{i}q_{i}}{4\pi\epsilon_{0}\epsilon_{\mathrm{r}}r_{ij}}\ .
\end{equation*}
$q_{i}$ is the partial charge of the $i$-th particle; $\epsilon_{0}$ is the permittivity
in vacuum and $\epsilon_{\mathrm{r}}$ is the relative dielectric constant. For the
standard MARTINI model $\epsilon_{\mathrm{r}}=15$, while for the polarizable model
used here $\epsilon_{\mathrm{r}}=2.5$. Non-bonded interactions are corrected separately
to zero using the shift function \cite{shift}
\begin{gather*}
S(r_{ij})=\\
\left\{\begin{array}{ll}
C & r_{ij}\leq r_{\mathrm{shift}}\\
\frac{A}{3}(r_{ij}-r_{\mathrm{shift}})^{3}+\frac{B}{4}(r_{ij}-r_{\mathrm{shift}})^{4}
+C & r_{\mathrm{shift}}<r_{ij}<r_{\mathrm{c}}\ ,\\
\end{array}
\right.
\end{gather*}
where $r_{\mathrm{shift}}$ is the inner cutoff distance and $r_{\mathrm{c}}$ is the outer
cutoff distance. The constants $A$, $B$ and $C$ are determined from boundary 
conditions. Intra-particle interactions take into account bond and angle interactions.
Bonded particles interact via the harmonic potential
\begin{equation*}
U_{\mathrm{b}}(r_{ij})=K_{\mathrm{b}}(r_{ij}-r_{0})^{2}\ ,
\end{equation*}
where $r_{0}$ is the equilibrium distance and $K_{\mathrm{b}}$ the energy constant.
The standard MARTINI values for these parameters are $r_{0}=4.7$ [\AA] and
$K_{\mathrm{b}}=1.5$ [kcal/mol/\AA$^{2}$]. Angle interactions are described by the
potential 
\begin{equation*}
U(\theta_{ijk})=K_{\mathrm{a}}[\cos(\theta_{ijk})-\cos(\theta_{0})]^{2}\ .
\end{equation*}
$\theta_{ijk}$ is the angle formed by triplets of bonded particles; $\theta_{0}$
is the equilibrium angle; $K_{\mathrm{a}}$ is the coupling constant, for which
the standard MARTINI value is $K_{\mathrm{a}}=3.0$ [kcal/mol].

\textit{Water}.~A CG bead of polarizable MARTINI water \cite{martini2} consists 
of three particles $W_{0}$, $W_{\pm}$ of equal mass $24$ [g/mol] and of partial 
charge $q_{0}=0$ [e] and $q_{\pm}=\pm 0.46$ [e], respectively. The charged particles 
are bonded to the neutral one and the bond length is kept fixed at $1.4$ [\AA].
The angle degree of freedom is modeled as a harmonic oscillator of
elastic constant $K_{\mathrm{a}}^{\mathrm{W}}=0.5019$ [kcal/mol] and
equilibrium angle of $0^{\circ}$. Lennard-Jones forces are
considered only among the central particles $W_{0}$ with interaction
parameters $\varepsilon_{\mathrm{WW}}=0.956$ [kcal/mol] and 
$\sigma_{\mathrm{WW}}=4.7$ [\AA]. With the charged 
particles $W_{\pm}$, the polarization is treated explicitly and the relative
dielectric constant is thus set to $\epsilon_{\mathrm{r}}=2.5$ 
(cf.~Refs.~\cite{martini1,martini2}). Furthermore, 
Coulomb and van der Waals forces among particles belonging to the same 
water bead are omitted.

\textit{Graphene}.~The graphene is parameterized by placing at the 
center of every ring of carbon atoms one interaction site. As
a consequence, every bead has mass $24$ [g/mol] because representing
on average two atoms. The interaction parameter $\varepsilon_{\mathrm{CW}}$
between the beads of water ($W_{0}$) and graphene is fixed from the
wetting properties of pure water. We choose the value of $\varepsilon_{\mathrm{CW}}$
that reproduces the result for the contact angle in the macroscopic
limit reported in Ref.~\cite{wang}. The parameter $\sigma_{\mathrm{CW}}$ 
is set equal to $\sigma_{\mathrm{CW}}=6.24$ [\AA]. This value yields approximately 
the equilibrium distance of $7$ [\AA]. The solid phase is 
obtained by two parallel planes of CG graphene. The two planes are 
separated by $3.4$ [\AA] with the lower one translated by the vector $(l/2,\sqrt{3}l/2)$
with respect to the upper one. Here $l$ is the length of C-C bonds
in the all-atom case. This arrangement is intended to reproduce the 
structure of a graphite crystal.

\textit{Surfactants}.~The atomic structure of the surfactants is reduced
according to the guidelines detailed in Ref.~\cite{martini1}.
In order to compare our work to previous similar studies \cite{sanders}, we also
consider solutions containing the non-ionic surfactant C$_{8}$E$_{4}$ (tetraethylene glycol octyl ether). 
We adopt the mappings C$_{1}$-C$_{1}$-P$_{5}$-P$_{5}$ and  C$_{1}$-C$_{1}$-Na-Na-P$_{4}$ proposed
in Ref.~\cite{sanders}. The first representation will be referenced as 
C$_{8}$E$_{4}$-P$_{5}$ surfactant while the latter as C$_{8}$E$_{4}$-NaNaP$_{4}$ 
\cite{sanders}. Both molecules are linear with every equilibrium bond length 
equal to $r_{0}=4.7$ [\AA]; all equilibrium angles are of course 
$\theta_{0}=180^{\circ}$. The energy constants are also the standard MARTINI
values \cite{martini1}, that is, $K_{\mathrm{b}}=1.5$ [kcal/mol/\AA$^{2}$] and 
$K_{\mathrm{a}}=3.0$ [kcal/mol]. For the mass of the beads, we use in this case
$m_{\mathrm{P}_{4}}=72$ [g/mol], $m_{\mathrm{Na}}=56$ [g/mol] and $m_{\mathrm{C}_{1}}=56$ [g/mol]. 
The LJ parameters are readily obtained from Ref.~\cite{martini1} 
with all equilibrium distances set to $4.7$ [\AA]. The wetting properties of
graphitic surfaces are studied in relation to five non-ionic surfactants, long-chain and T-shaped. 
The first surfactant will be referred to as L1. 
Its molecule is linear and it is applied the MARTINI representation $($C$_{1})_{3}($P$_{4})_{10}$. 
The mass is again $m_{\mathrm{C}_{1}}=56$ [g/mol] for the tail beads, while we use 
$m_{\mathrm{P_{4}}}=44$ [g/mol] for the hydrophilic head. 
The second surfactant has representation $($C$_{1})_{3}($P$_{4})_{20}$ and will be designated by L2.
All topology and interaction parameters are the same as for the surfactant L1.
The third linear surfactant L3 is mapped into $($C$_{1})_{4}($P$_{4})_{30}$. 
All the parameters are the same as for the  surfactant L1. Regarding the T-shaped surfactants,
we distinguish the T1 and T2 topologies. These surfactants differ by the length
of the hydrophilic headgroup. The surfactant T1 has a linear chain made of ten
beads P$_{4}$, while the surfactant T2 counts twenty beads P$_{4}$ arranged linearly.
In both cases, the hydrophobic tail consists of three aligned beads C$_{1}$ and in the
middle is attached the hydrophilic chain. For the angle C$_{1}$-C$_{1}$-P$_{4}$
(vertex in the central bead), the equilibrium value is set to $90^{\circ}$, and
the standard potential and interaction parameter of the MARTINI
force field \cite{martini1} are used. All the other parameters are the same as for the linear
surfactants. The hydrophobic tail of the surfactant T3 is made of five C$_{1}$ beads.
At the central bead is attached the linear, hydrophilic head of length twenty P$_{4}$ units.
All the other settings for T-shaped surfactants apply also in this case.

%======================================================================

\section{Simulations}\label{sec:sim}

All simulations are performed with the molecular dynamics code LAMMPS
\cite{pppm,code,parallel}. Newton's equations of motion are integrated with
the Verlet algorithm using a Nos\a'e-Hoover scheme in the specified ensemble; 
the timestep size is of $20$ [fs]. 
Non-bonded interactions are cut off at $r_{\mathrm{c}}=12$ 
[\AA] and computed with the potential lj/gromacs/coul/gromacs \cite{shift}.
The LJ potential is shifted from $r_{\mathrm{shift}}=9$ [\AA]
to $r_{\mathrm{c}}$, while the electrostatic contribution is shifted 
from $r_{\mathrm{shift}}=0$ [\AA] to $r_{\mathrm{c}}$. Long-range interactions
are not taken into account. The bond length between $W_{0}$ and $W_{\pm}$
particles is maintained rigid with the SHAKE algorithm \cite{shake}. 
Three-body interactions of surfactants are computed with the cosine/squared 
potential. The neighbor list is rebuilt at most every $5$ timesteps. It is not
necessary to divide the bond and angle coefficients given here by $2$. Then, 
all simulation times must be meant as actual \cite{martini2}. 
Unless specified otherwise, these parameters remain unchanged in the various
simulations. The motivations for these settings can be found 
in Refs.~\cite{martini1,martini2}.
%======================================================================

\section{Results and discussion}\label{sec:results}

\begin{figure}[t]
\includegraphics[width=8.5cm]{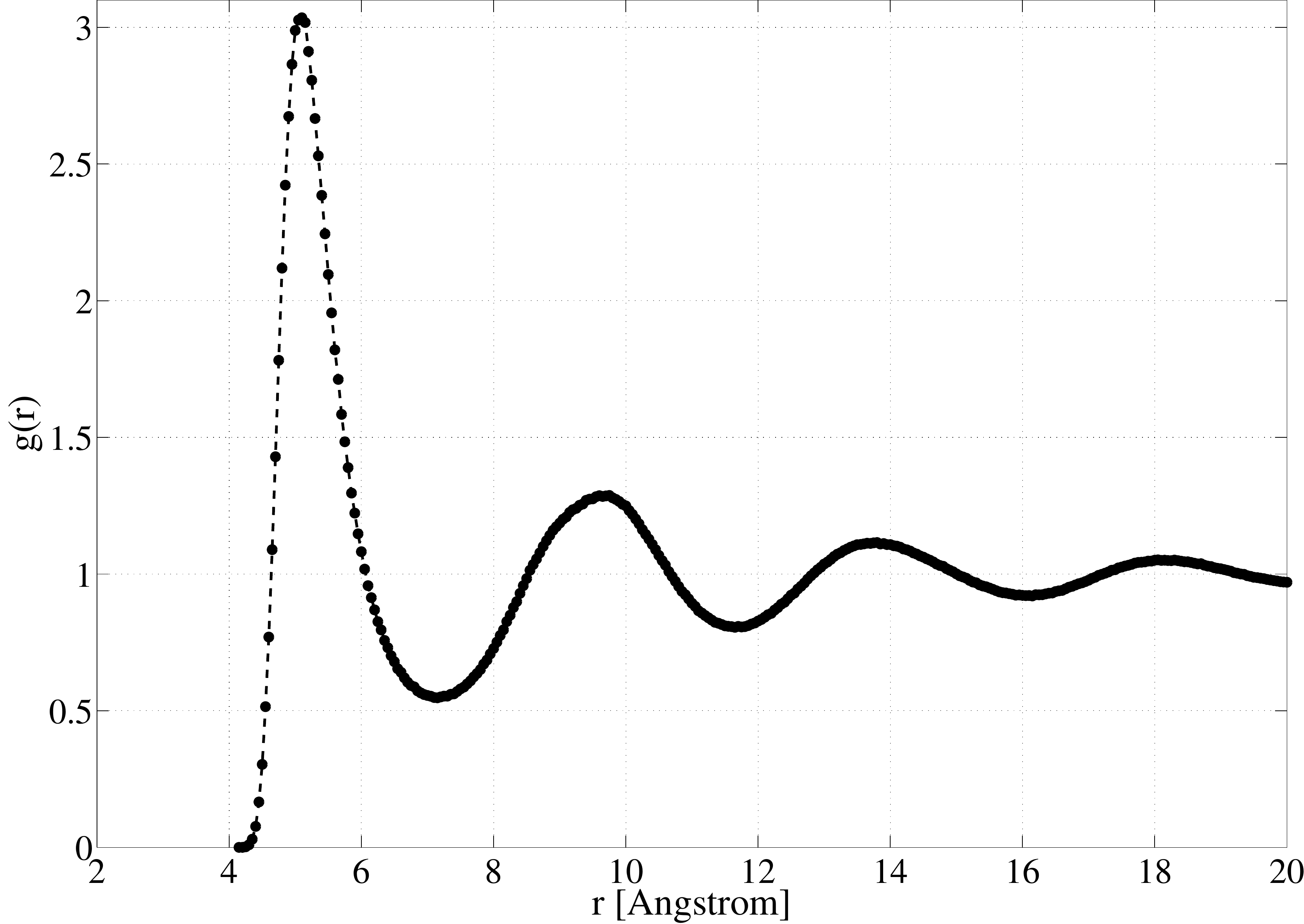}
\caption{
$W_{0}$-$W_{0}$ RDF as obtained from the NPT
dynamics for the polarizable MARTINI water model.
\label{fig:rdf}}
\end{figure}

\textit{Bulk properties of CG water}.~As starting configuration,
$1'000$ beads of water are arranged with the $W_{0}$ particles on the
vertices of a simple cubic lattice. The side of the unit cell is $4.5$ [\AA].
The system is let evolve for $100$ [ns] at NPT conditions. 
The time of evolution is long in order to verify that we do not incur in  freezing.
The target temperature and pressure are $T=298$ [K] and $P=1$ [atm]. The system
is studied by recording $1'000$ evenly-spaced frames over the course of the
last $50$ [ns]. Table \ref{tab:bulk_w}
lists some average bulk properties of the cubic box of CG water. 
The $W_{0}$-$W_{0}$ radial distribution function, $g(r)$, is computed by means
of the formula
\begin{equation*}
\frac{N}{V}g(r)4\pi r^{2}\Delta r=S(r)\ .
\end{equation*}
$N$ is the number of water beads and $V$ the volume of the simulation domain. 
$S(r)$ counts the average number of $W_{0}$ particles falling
in a shell of width $\Delta r=0.05$ [\AA] centered around a given $W_{0}$
particle at distance $r$. The radial distribution function (RDF) is shown in 
Fig.~\ref{fig:rdf}. In general, our results compare well with those of
the original work for the present model of CG water 
\cite{martini2}. The standard deviation of the volume 
of the simulation domain is $0.78\%$ of the average value. For this reason, we 
safely replicate the last configuration of this NPT simulation and extract the 
asseblies for any subsequent dynamics containing pure CG water.
%=======================================
\begin{table}[t]
\begin{ruledtabular}
\begin{tabular}{cccc}
$L$ [\AA] & $V$ [\AA$^{3}$] & $\rho$ [g/\AA$^{3}$] & $d$ [molecules/\AA$^{3}$]\\
\hline
$48.858$ & $116'629$  & $1.0255$ & $0.0086$
\end{tabular}
\end{ruledtabular}
\caption{\label{tab:bulk_w}
Average characteristics of the cubic simulation box during the NPT
dynamics, side length $L$ and volume $V$, along with the bulk properties
of CG water $\rho$ and $d$, mass and particle densities 
respectively.}
\end{table}
%=======================================
%=======================================
%=======================================
%=======================================
\begin{figure}[t]
\includegraphics[width=8.5cm]{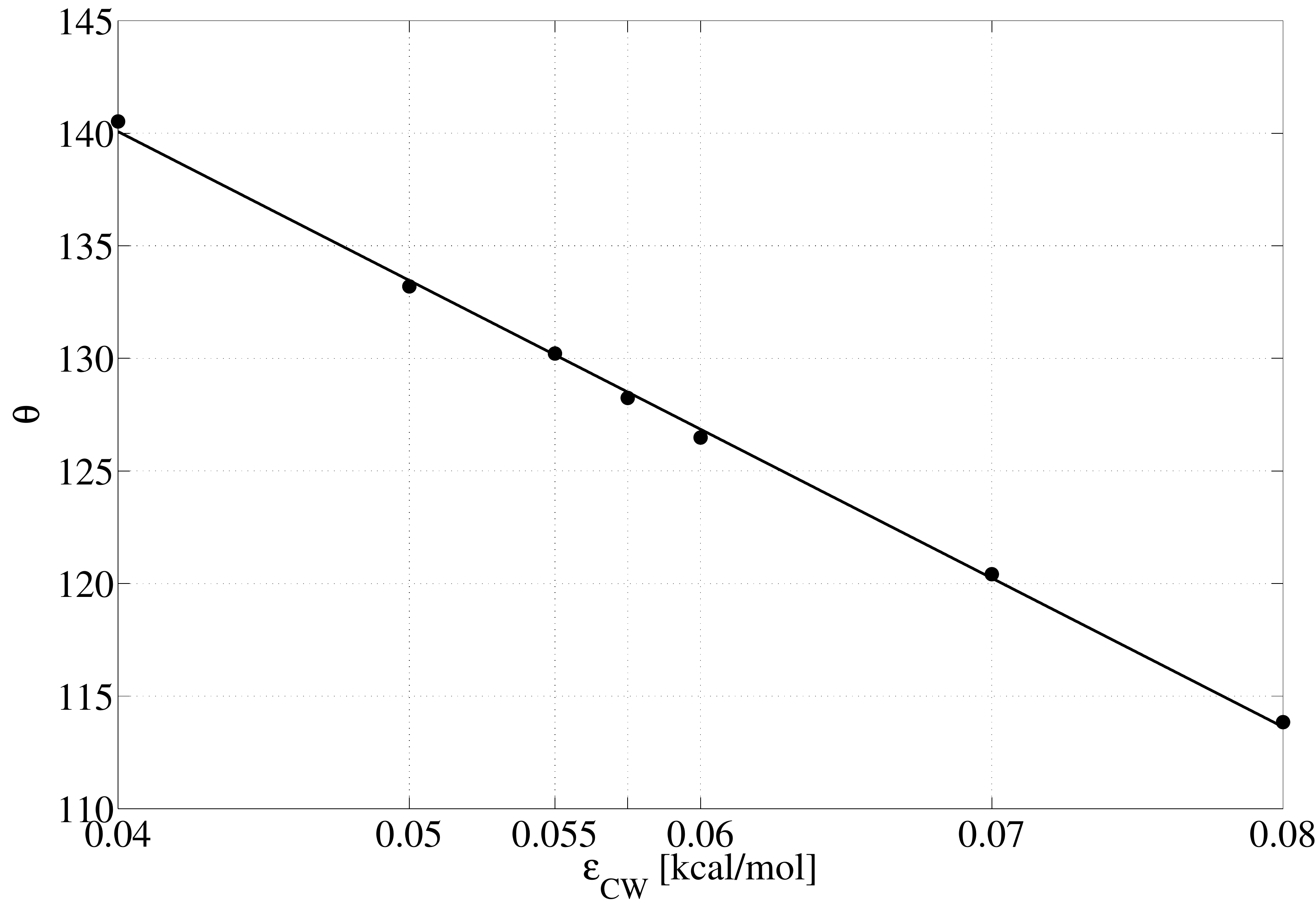}
\caption{
Contact angle as a function of the strength parameter. Matching with the experimental 
result \cite{wang} occurs approximately for $\varepsilon_{\mathrm{CW}}=0.0575$ [kcal/mol].
\label{fig:e2temp}}
\end{figure}
%=======================================

\textit{Spherical droplets}.~As a matter of calibration of the 
interaction parameter $\varepsilon_{\mathrm{CW}}$, droplets 
of pure CG water spreading on the CG graphene are 
first considered. The interaction strength $\varepsilon_{\mathrm{CW}}$ is 
varied in the range $0.04-0.08$ [kcal/mol]. Every initial configuration is 
composed by a hemispherical droplet of $17'927$ beads centered above the upper-lying 
plane of graphene at a distance of $7$ [\AA]. The radius of the hemisphere is $100$
[\AA]. The planes of graphene are squared with side length $320$ [\AA]. 
The boundary conditions are periodic. Every system is evolved in the canonical
ensemble for $60$ [ns] at $T=298$ [K]. 
The systems are studied from $1'000$ frames recorded during the last
$40$ [ns]. These dynamics are shorter because for long times of evolution the
droplets systematically exhibit excessive stratification. The onset of this
phenomenon depends on several parameters, first of all the equilibrium distance
of the liquid interface from graphene. Presumably, this shortcoming might be avoided 
with an involved work of refinement of the force field. We do not insist on this aspect 
because we expect that the presence of surfactant molecules would reduce the formation 
of layers of solvent particles. Furthermore, higher temperatures could
retard this phenomenon because of faster kinetics. Then, the behavior at room temperature 
could be extrapolated. For all simulations presented in this Article, stratification never 
occured and, as we shall see, our outcomes are still in reasonable agreement 
with the atomistic results and the general theoretical predictions. 
%=======================================
\begin{figure}[t]
\includegraphics[width=4cm]{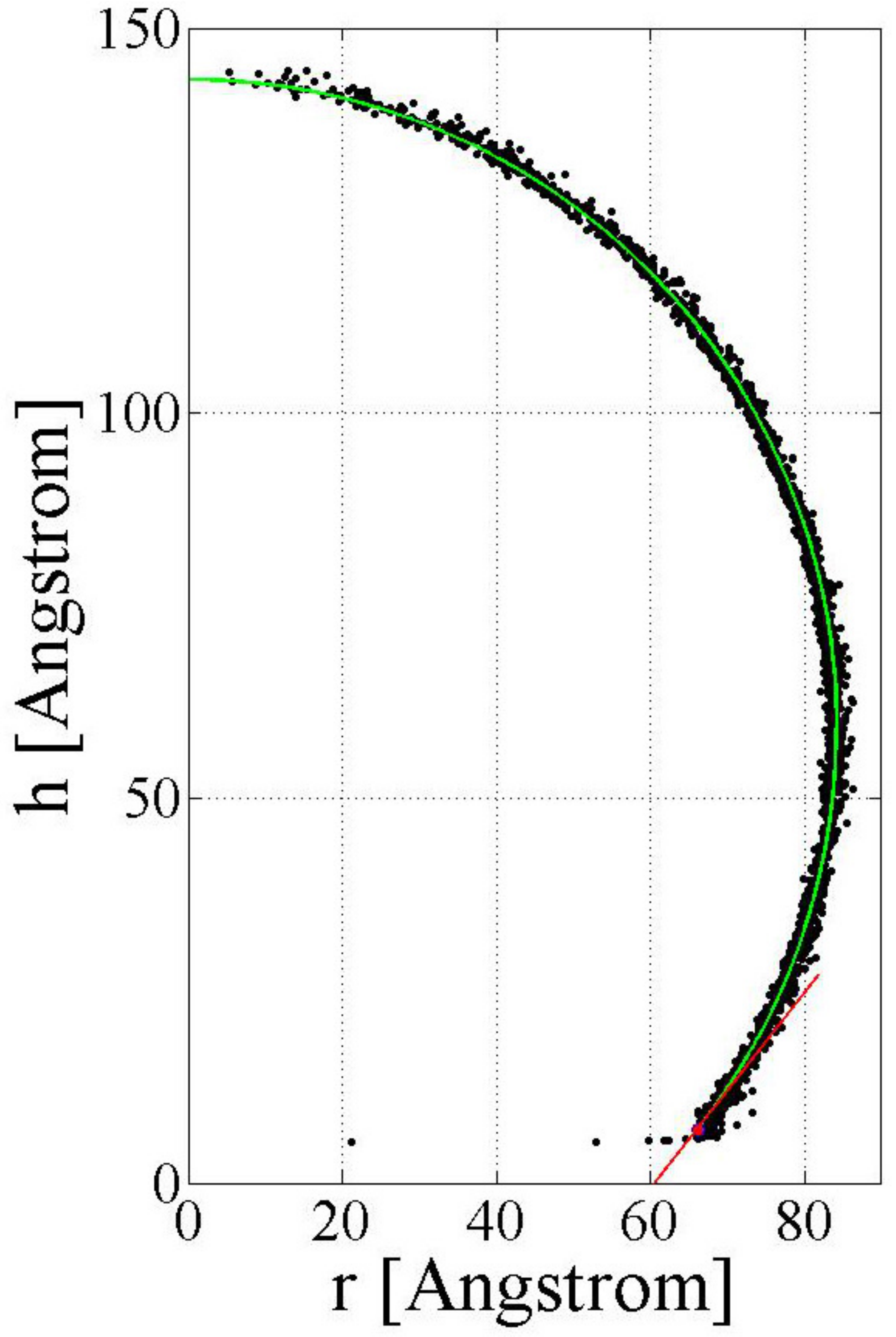}
\caption{
(Color online) Profile of the spherical droplet for $\varepsilon_{\mathrm{CW}}=0.0575$ [kcal/mol]. 
The data are analyzed following the method presented in Refs.~\cite{epsilon,binning,werder}. 
The green line is a fit to the data under the assumption of a spherical shape of the droplet. 
The tangent at the contact line, where the solid, liquid and vapor phases meet, is in red and 
the contact angle is measured to be $128.2^{\circ}$. 
\label{fig:profile}}
\end{figure}
%=======================================
Figure \ref{fig:e2temp} shows the dependence of the contact angle on the parameter
$\varepsilon_{\mathrm{CW}}$. These data clearly satisfy a linear relation as 
in the atomistic case \cite{epsilon}. For $\varepsilon_{\mathrm{CW}}=0.0575$ [kcal/mol],
it is found a contact angle of $128.2^{\circ}$. This result reproduces the
experimental value reported in Ref.~\cite{wang}. The profile of the droplet
in this case is shown in Fig.~\ref{fig:profile}. In principle, other coarse-grained 
force fields are available for such
investigations \cite{shinoda1,shinoda2}. For example, in Ref.~\cite{sdk_droplet} 
the wetting properties of pure water on a flat surface were investigated by taking 
a LJ function of type 6-3 for the interaction between the liquid and solid phases. 
Using the same CG water, preliminary results indicate that the process of stratification
on a molecular graphitic substrate occurs quite fast with the liquid-solid interactions 
described by a LJ potential of type 12-6. Ultimately, the assessment of other 
coarse-grained force fields for wetting studies warrants a detailed discussion, 
which is beyond the scope of the present study.

%=======================================
\begin{table}[t]
\begin{tabular}{c|c|c}
\hline\hline
$r$ [\AA] & no.~water beads & $\theta$ [$^\circ$]\\
\hline
$80$  & $25'933$ & $128.0^{\circ}$ \\
$90$  & $32'742$ & $128.7^{\circ}$ \\
$100$ & $40'430$ & $129.0^{\circ}$ \\
$110$ & $48'800$ & $129.9^{\circ}$ \\
$120$ & $58'173$ & $130.4^{\circ}$ \\
$130$ & $68'255$ & $130.4^{\circ}$ \\
\hline\hline
\end{tabular}
\caption{\label{tab:theta}
Contact angle as a function of the initial radius of cylinders. The
variations of $\theta$ indicate that the macroscopic regime is
reached for radii $\sim 130$ [\AA].}
\end{table}
%=======================================
%=======================================
\begin{table}[t]
\begin{tabular}{c|c|c|c}
\hline\hline
$T$ [K] & $d$ [molecules/\AA$^{3}$] & $r_{\mathrm{f}}$ [\AA] & $\theta$ [$^\circ$]\\
\hline
$298$ & $0.0076$ & $100.3$ & $130.4^{\circ}$ \\
$305$ & $0.0075$ & $101.8$ & $129.3^{\circ}$ \\
$310$ & $0.0075$ & $101.1$ & $130.6^{\circ}$ \\
$315$ & $0.0075$ & $101.3$ & $130.8^{\circ}$ \\
$320$ & $0.0074$ & $101.4$ & $131.4^{\circ}$ \\
$350$ & $0.0071$ & $102.0$ & $132.9^{\circ}$ \\
$400$ & $0.0065$ & $104.1$ & $136.2^{\circ}$ \\
\hline\hline
\end{tabular}
\caption{\label{tab:temp}
Variations from temperature rise for the cylinder of initial radius $130$ [\AA]: 
particle density $d$, final radius $r_{\mathrm{f}}$ and contact angle $\theta$.}
\end{table}
%=======================================

\textit{Cylindrical droplets}.~We now want to investigate the dependence of the contact angle 
on size and temperature. To achieve this, six hemicylindrical droplets of radii from $80$ to $130$ [\AA]
are considered. The $x$ side of the simulation domain is always $300$ [\AA],
corresponding to the height of every hemicylinder. The $y$ side 
is always $120$ [\AA] larger than the diameter of the droplets. In the initial configuration,
the droplets are centered in the $xy$ plane. All other simulation settings are the 
same employed for spherical droplets. In Tab.~\ref{tab:theta} are listed the contact angles
measured from the profiles of cylinders with increasing radii. The contact angles vary
only of a few degrees and thus we conclude that size effects are negligible with radii $\sim 130$ [\AA]. 
It is interesting to remark that the contact angle for cylinders is slightly larger
than that for the spherical droplet, as observed for atomistic systems approaching
the macroscopic regime \cite{epsilon}. For the largest cylinder, different temperatures
are considered. 
The results of Tab.~\ref{tab:temp} suggests that variations over a range of $100$ [K]
have no drastic effect. The results reported here for the temperature dependence of
contact angle are in line with those for atomistic systems \cite{contact}.
In the following, cylindrical droplets are preferred to spherical ones because the
simulations turn out to be accelerated \cite{york}. More precisely, the CPU timings for
the cylindrical droplet of radius $80$ [\AA] are comparable with those of
spherical droplets, of radius $100$ [\AA]. The former system contains $25'933$ water beads
while the latter $17'927$. The reason for the speed-up is that parallelization based on
spatial decomposition \cite{parallel} is more efficient for cylinders because the 
projection of the particles on the $xy$ plane of the simulation domain gives a better coverage.

%=======================================
\begin{figure*}[t]
\includegraphics[width=8.5cm]{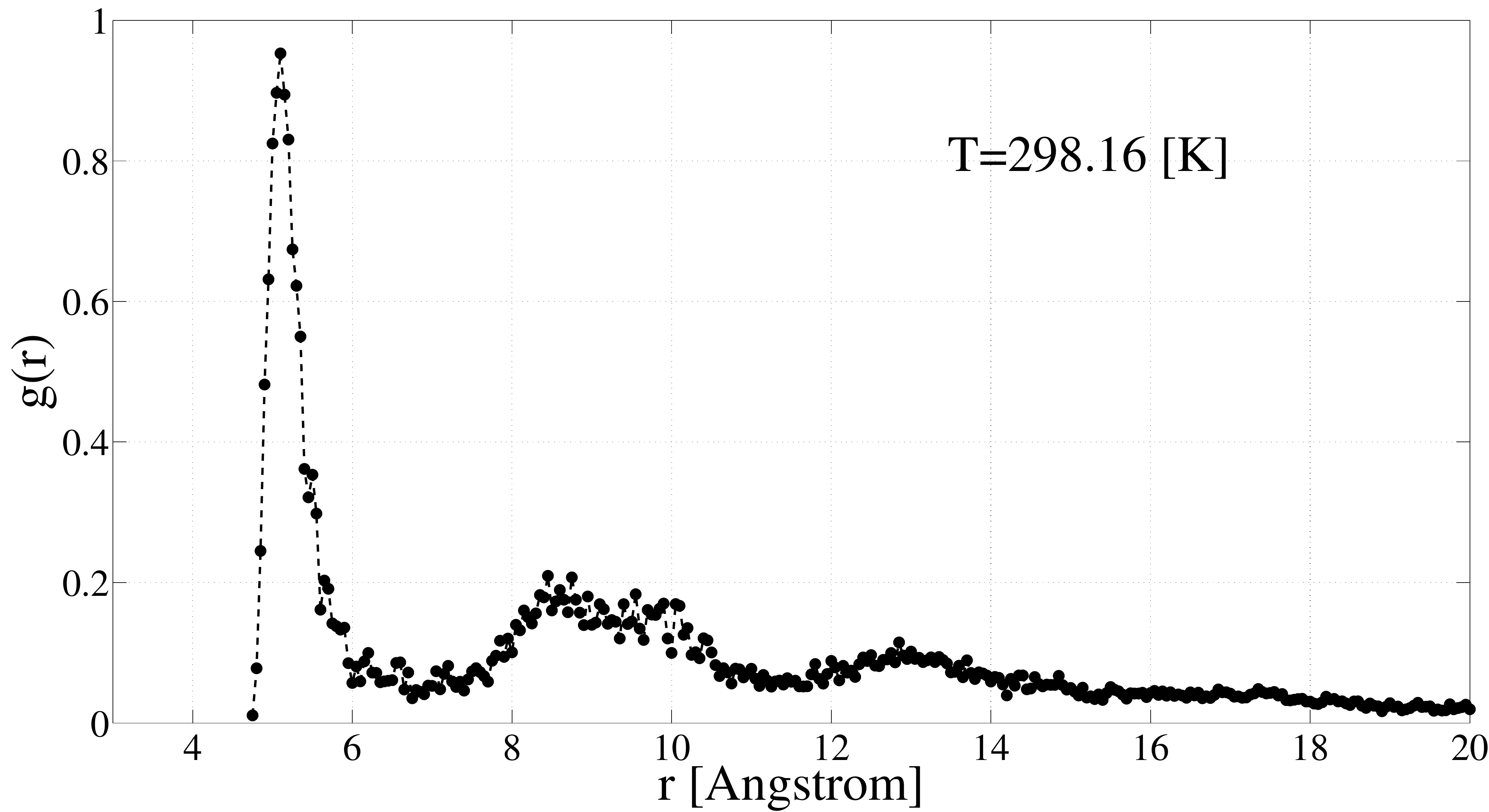}\hspace{0.5cm}
\includegraphics[width=8.5cm]{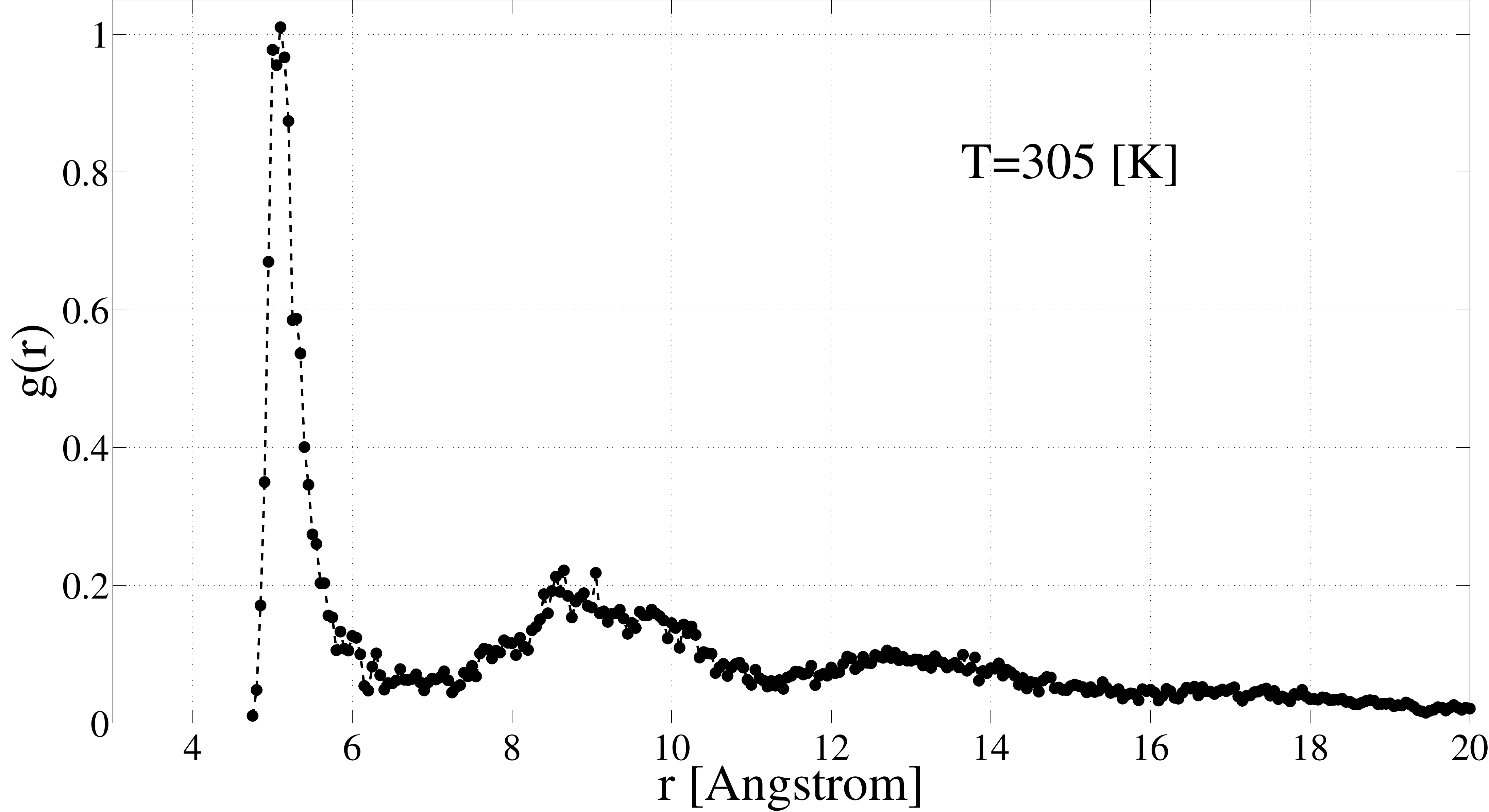}\\
\includegraphics[width=8.5cm]{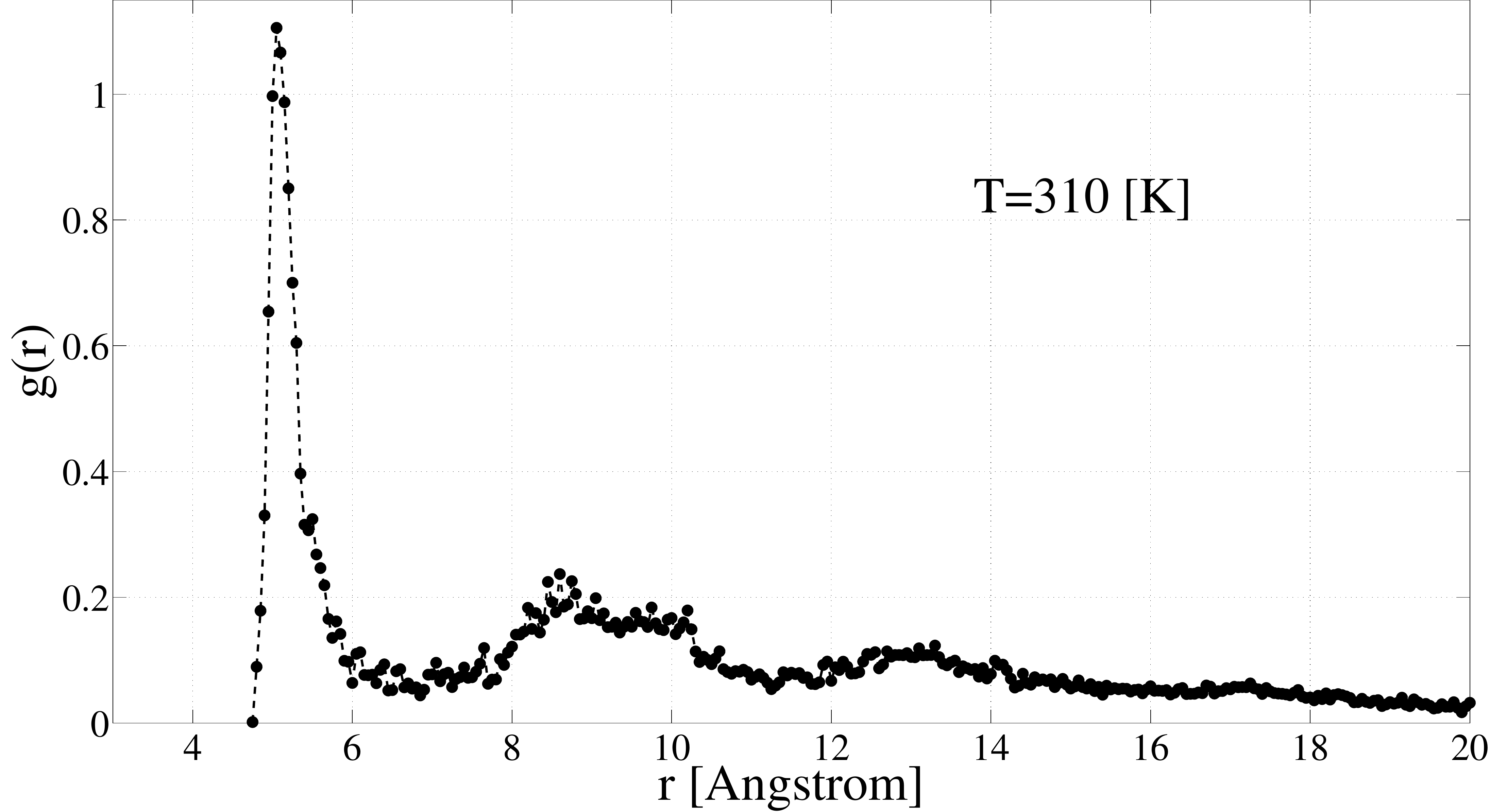}\hspace{0.5cm}
\includegraphics[width=8.5cm]{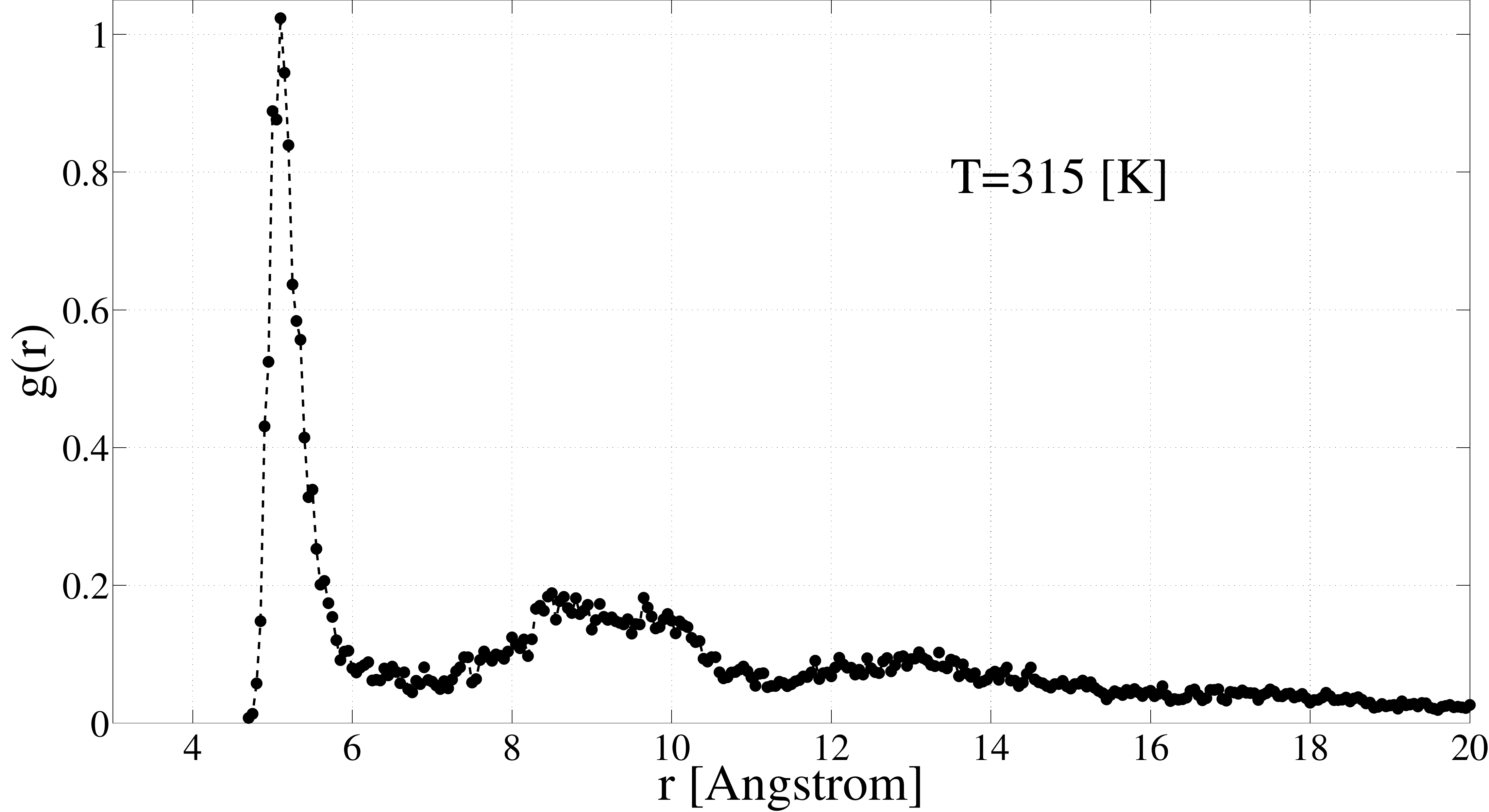}
\caption{
RDFs relative to the tail beads C$_{1}$ of the surfactant C$_{8}$E$_{4}$-NaNaP$_{4}$
from isobaric-isothermal dynamics. Average is taken over the last $100$ [ns] of the 
whole evolution.
\label{fig:rdf1}}
\end{figure*}
%=======================================
\begin{table*}[t]
\begin{tabular}{l|c|c|c|c|c|c|c|c}
\hline\hline
Mapping & $T$ [K] & Volume [\AA$^{3}$] & $C$ [M] & Average & Variance & Size range  & no.~monomers & CMC [M]\\
\hline
\multirow{4}{*}{C$_{8}$E$_{4}$-NaNaP$_{4}$} & $298$ & $1'845'116$ & $0.158$ &  $17.5$ & $81.8$  & $5$-$38$  & $0.0$ & $-$\\
                                          & $305$ & $1'846'956$ & $0.158$ &  $25.0$ & $244.9$ & $7$-$53$  & $0.0$ & $-$\\
                                          & $310$ & $1'847'152$ & $0.158$ &  $25.0$ & $150.1$ & $8$-$42$  & $0.0$ & $-$\\
                                          & $315$ & $1'847'130$ & $0.158$ &  $25.0$ & $305.0$ & $11$-$63$ & $0.0$ & $-$\\
\hline
C$_{8}$E$_{4}$-P$_{5}$                      & $350$ & $1'879'769$ & $0.154$ & $6.0$ & $21.1$ & $1$-$18$  & $5.0$ & $4.4\times 10^{-3}$\\
\hline\hline
\end{tabular}
\caption{\label{tab:stat_surf}
Micelle size statistics over the last $100$ [ns] for the simulations of
surfactant C$_{8}$E$_{4}$. With volume we intend the average value
of the simulation domain; its standard deviation is always below $0.1\%$. The symbol $C$ designates the concentration.}
\end{table*}
%=======================================

\begin{figure}[t]
\includegraphics[width=8.5cm]{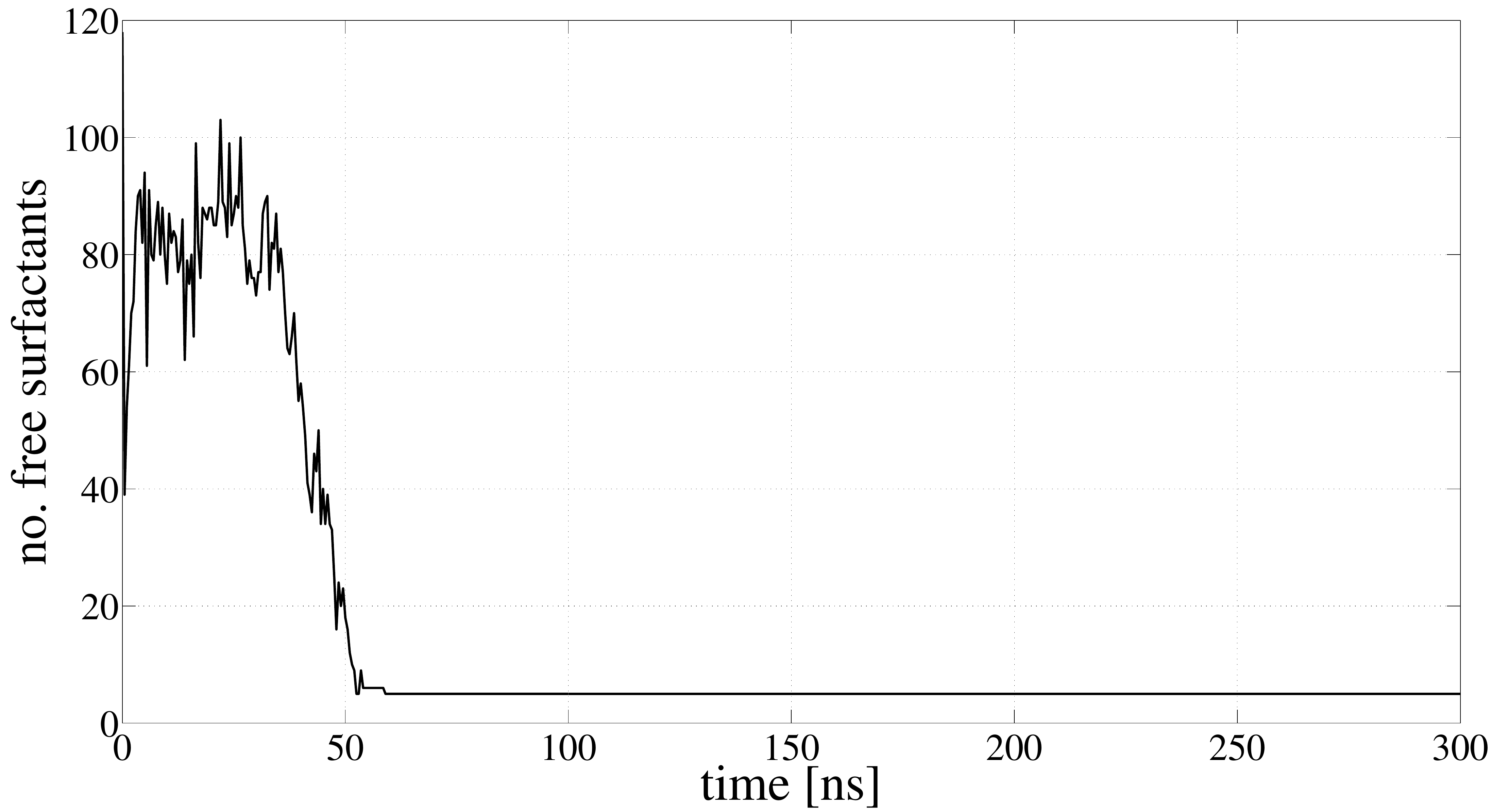}
\caption{Evolution of the number of free surfactants for the C$_{8}$E$_{4}$-P$_{5}$
solution at $350$ [K] (see Tab.~\ref{tab:stat_surf}).
\label{fig:free_p5}}
\end{figure}

\begin{figure*}[t]
\includegraphics[width=8.5cm]{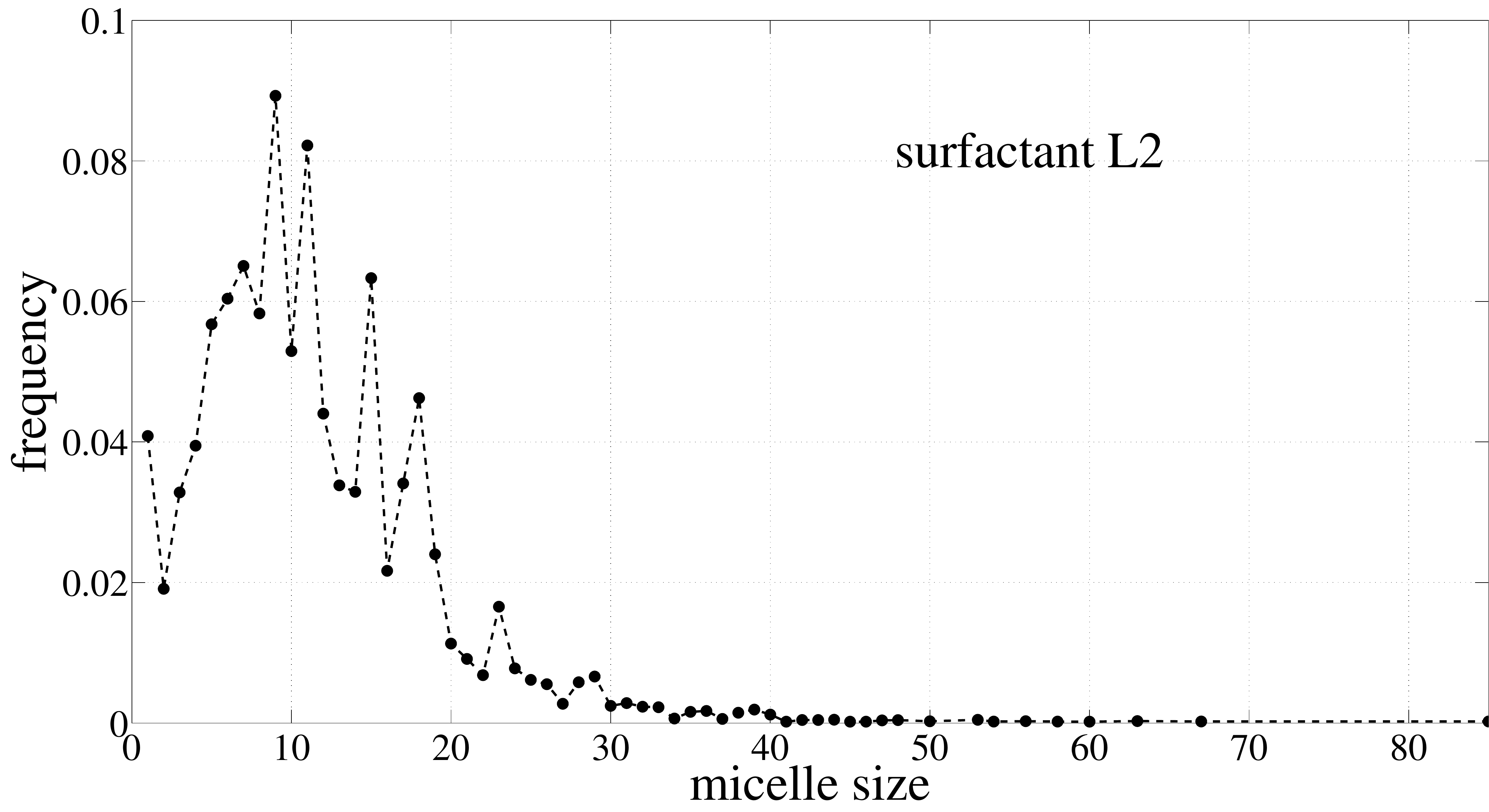}\hspace{0.5cm}
\includegraphics[width=8.5cm]{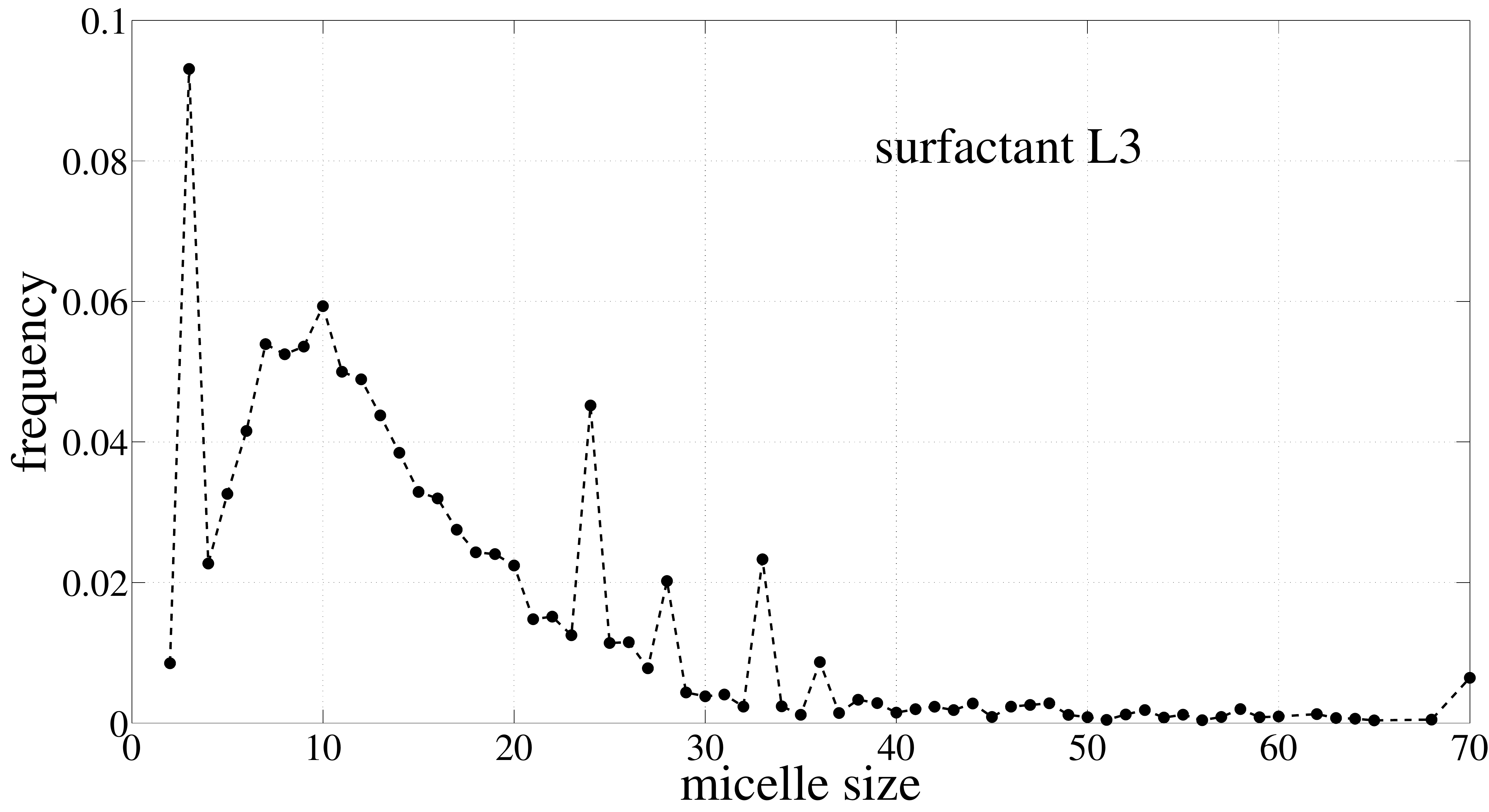}\\
\includegraphics[width=8.5cm]{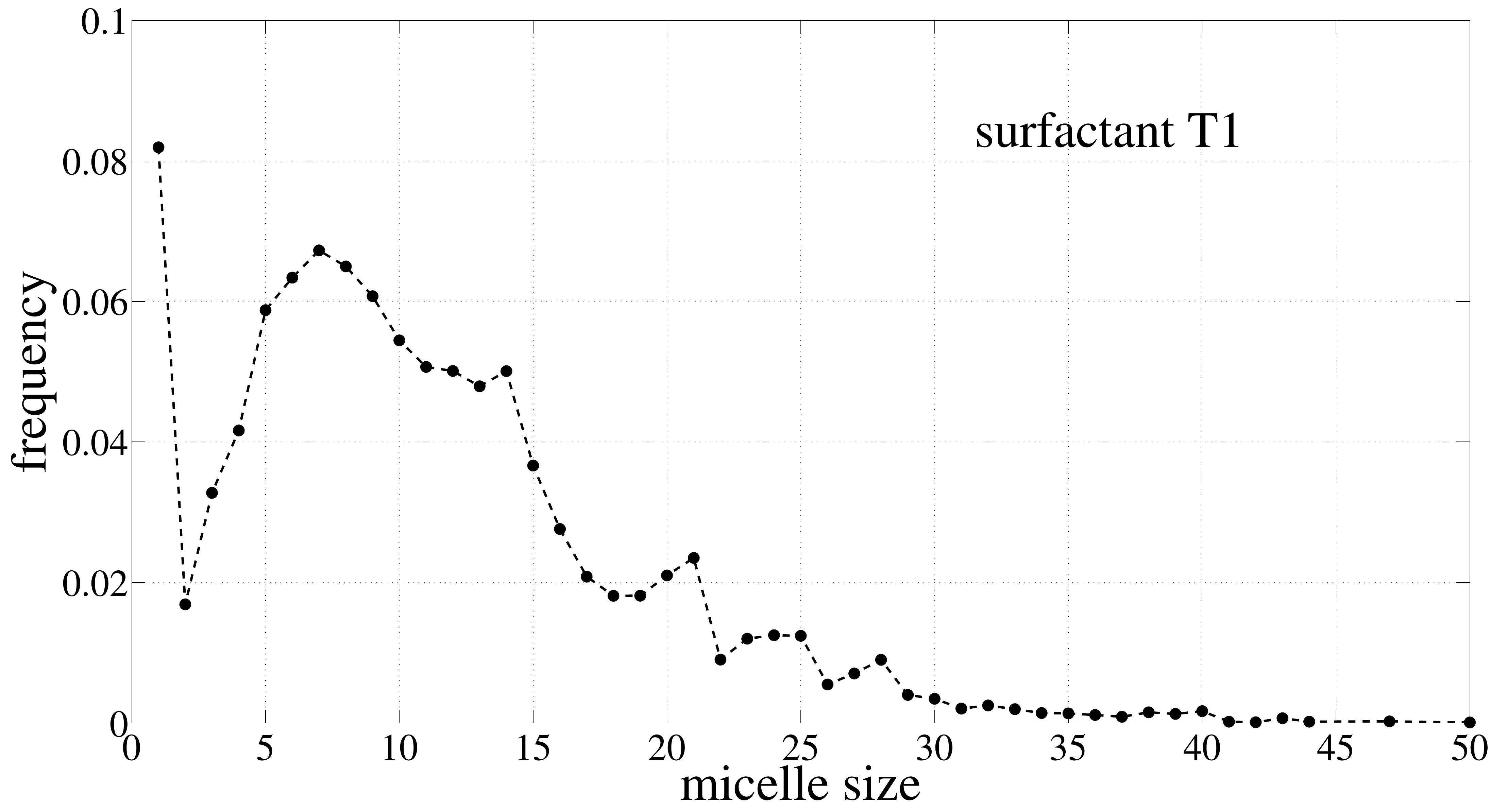}\hspace{0.5cm}
\includegraphics[width=8.5cm]{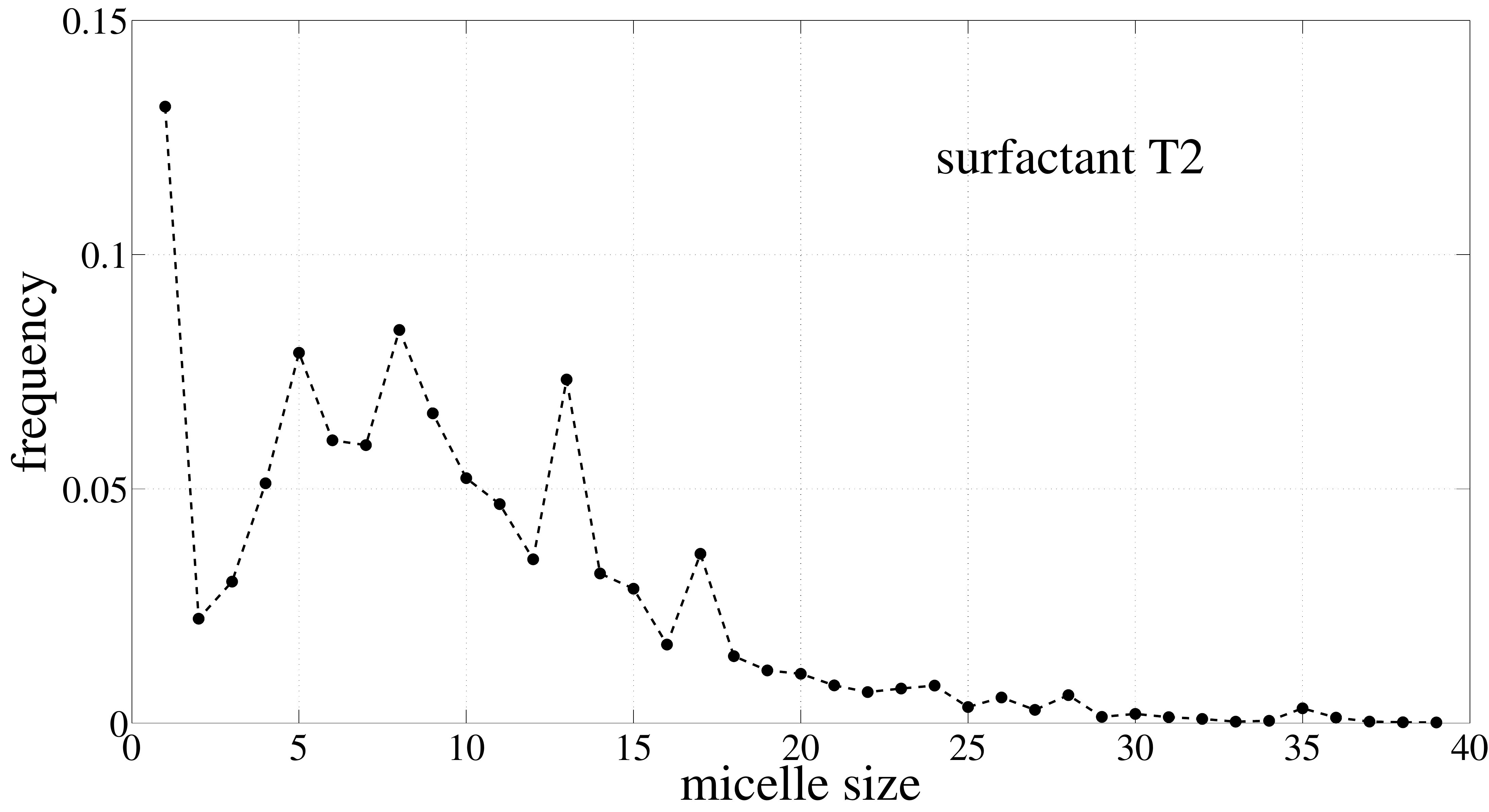}
\caption{
Micelle size distributions as resulting from the simulations of equilibration
for two long-chain, linear surfactants and the T-shaped surfactants at $8$ wt$\%$ concentration. Top: For the
L2 surfactant the distribution is peaked around the size of $10$, while for the longer surfactant L3 the more 
frequent micelle is of size $3$. Bottom: T-shaped surfactants have less micelle aggregates composed of 
fifteen or more surfactants. This occurrence has probability $0.26$ for the surfactant T1 and $0.18$ for
the surfactant T2. The same probabilities in the linear case are $0.33$ and $0.37$ for the surfactants
L2 and L3, respectively.
\label{fig:freq_mic}}
\end{figure*}

\textit{Bulk properties of surfactant solutions}.~Here we consider boxes of CG water 
containing surfactants. In the initial configuration, the surfactant molecules 
are arranged regularly between two slabs of water. Every simulation lasted for $300$ [ns] at NPT conditions. 
The target pressure is always $1$ [atm] while the temperature is adjusted
to different values. Data for analysis are recorded every $5$ [ns].
To start with, we want to compare the predictions of the MARTINI model with
polarizable water \cite{martini2} to those of the standard version \cite{martini1}. 
To this end, we first simulate $175$ surfactant molecules of C$_{8}$E$_{4}$-NaNaP$_{4}$ 
with $18'339$ water beads, as done in Ref.~\cite{sanders}. This
concentration leads to the formation of micelles \cite{sanders}. We look at the RDF between
C$_{1}$ beads belonging to different surfactant molecules in order to extract
the equilibrium distance between tail beads in the same micelle. From Fig.~\ref{fig:rdf1},
we see that there is a first peak around $5.1$ [\AA], while the first minimum
occurs at about $r_{\mathrm{min}}=7$ [\AA]. As a consequence, we shall assume that
the surfactants with all of their C$_{1}$ beads separated by more than $r_{\mathrm{min}}$
are free with respect to each other. The first pronounced peak, together with the fact that
the RDF tends to zero, indicate that
the tail beads are confined in relatively small regions of
the whole simulation domain. Indeed, for all temperatures it is found that
most of the surfactant molecules cluster into micelles, corroborating the results reported 
in Ref.~\cite{sanders}. Table \ref{tab:stat_surf} lists basic statistical indicators
relative to the size distribution of micelles. As the temperature increases, the average
size of micelles increases and there is clearly more dispersion around the mean value
(enhanced polydispersity). Furthermore, the size of micelles (aggregation number) tends
to increase with temperature, as expected for non-ionic surfactants \cite{book_surf}. In that
respect, it is interesting to remark that the first peak of the C$_{1}$-C$_{1}$ (tail beads)
RDF is more marked at higher temperatures (see Fig.~\ref{fig:rdf1}). For the other representation
of the surfactant C$_{8}$E$_{4}$ \cite{sanders} (see Sec.~\ref{sec:cg}), we also consider $175$ 
molecules with $18'339$ water beads at the temperature of $350$ [K]. The C$_{1}$-C$_{1}$ RDF for the
surfactant C$_{8}$E$_{4}$-P$_{5}$ is similar to those of Fig.~\ref{fig:rdf1} and the characteristic
distance $r_{\mathrm{min}}$ associated with micelle formation is $7$ [\AA] also in this case. 
For this mapping, the headgroup is more polar and the process of micelle formation is effectively 
weakened. Figure \ref{fig:free_p5} shows the number of free surfactants in the course of time. 
Table \ref{tab:stat_surf} presents some statistics for micelles. The critical micelle concentration 
(CMC) is the concentration at which micelles start forming. We calculate the CMC from the 
average number of free surfactants \cite{sanders}. In this way, it is understood that the 
removal of micelles leaves the surfactant solution at CMC. After $300$ [ns] of dynamics
there remain five free surfactant molecules (see Fig.~\ref{fig:free_p5} and Tab.~\ref{tab:stat_surf}):
it is thus predicted a CMC of $4.4\times 10^{-3}$ [M]. Now, the CMCs at two different
temperatures $T_{1}$ and $T_{2}$ are related by $\mathrm{CMC}(T_{2})=[\mathrm{CMC}(T_{1})]^{T_{1}/T_{2}}$ \cite{cmc,sanders}.
From the experimental value $\mathrm{CMC}(T_{1}=298\mathrm{ [K]})=8\times 10^{-3}$ [M] \cite{cmc_exp1,cmc_exp2}
it is obtained a CMC of $16\times 10^{-3}$ at $350$ [K]. It turns out that our result
underestimates this value but is closer than the finding within the standard MARTINI model of 
$35\times 10^{-3}$ [M] \cite{sanders}.
%=======================================
%=======================================
\begin{figure*}[b]
\includegraphics[width=8.5cm]{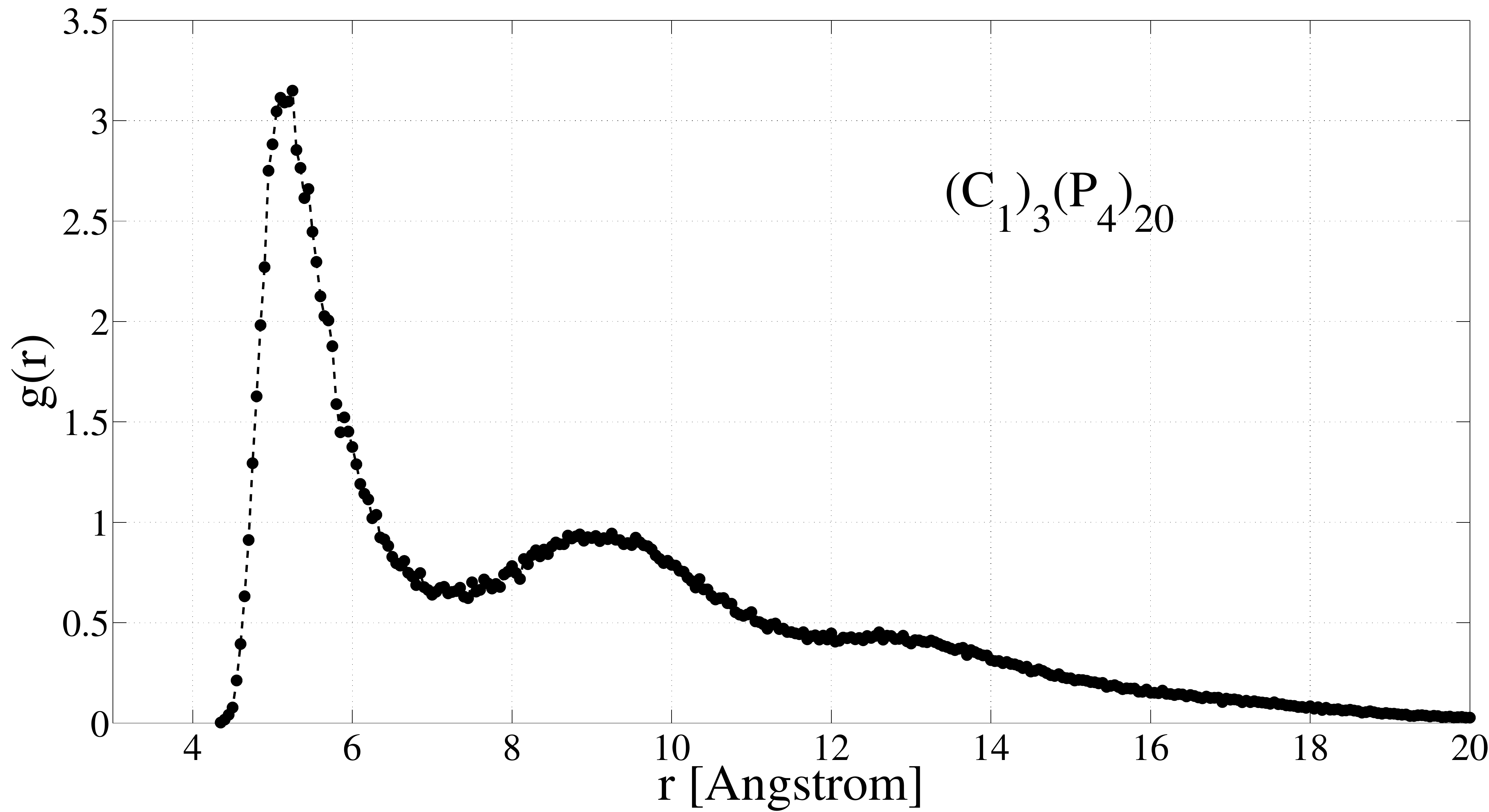}\hspace{0.5cm}
\includegraphics[width=8.5cm]{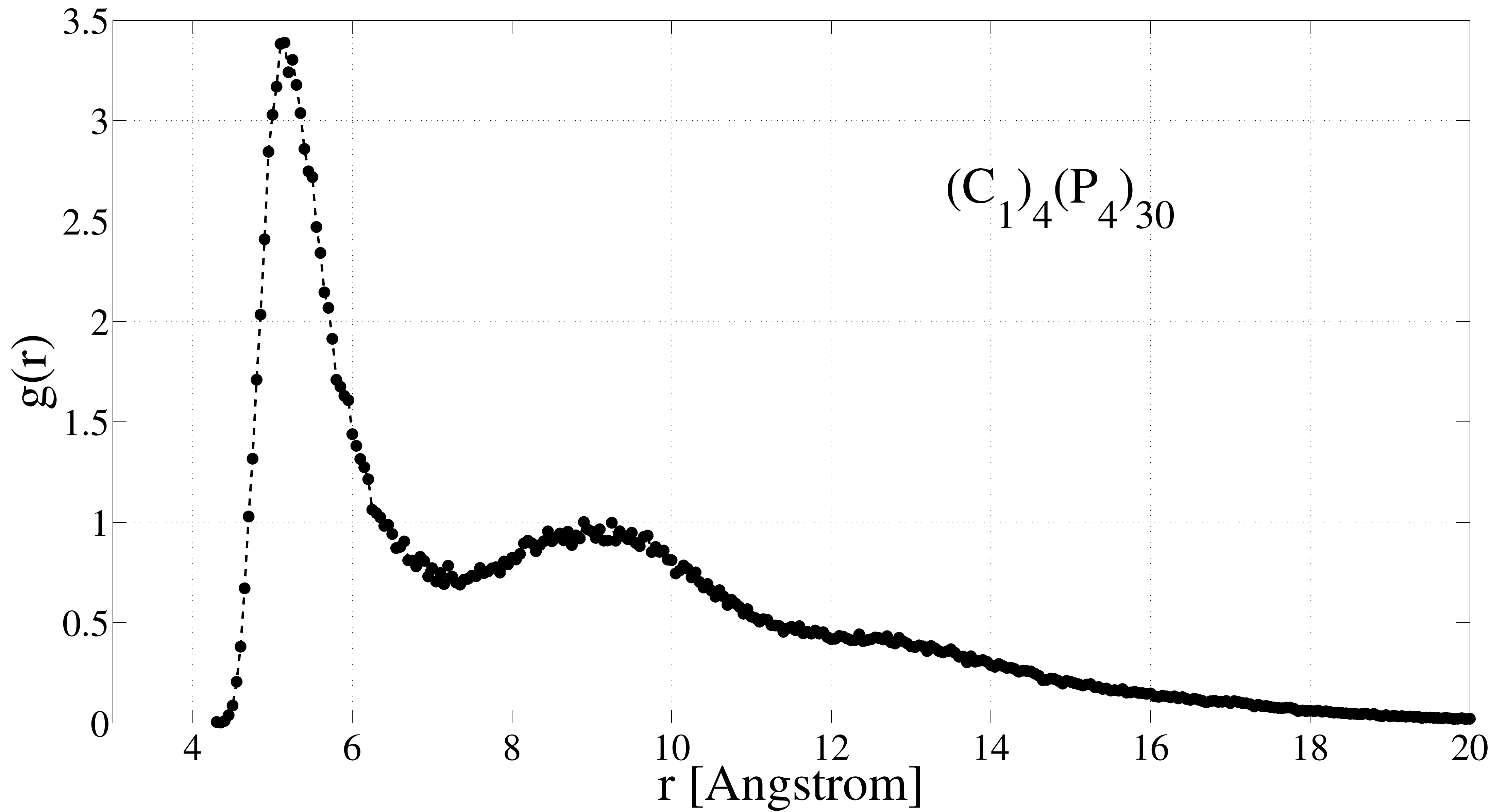}
\caption{
C$_{1}$-C$_{1}$ RDFs for the boxes with linear surfactants at $1$ wt$\%$; Left: surfactant L2; 
Right: surfactant L3. Similar results are obtained for the other equilibrated boxes with higher 
concentrations of surfactant molecules (cf.~Tab.~\ref{tab:stat_bip}).
\label{fig:rdf_bip}}
\end{figure*}
%=======================================
%=======================================
\begin{table*}[b]
\begin{tabular}{l|l|c|c|c|c|c|c|c|c}
\hline\hline
Surfactant & Mapping & Volume [\AA$^{3}$] & no.~surfactants & $C$ [M] - [wt$\%$]& Average & Variance & Size range & no.~monomers & CMC [M]\\
\hline
\multirow{2}{*}{L2} & \multirow{2}{*}{$($C$_{1})_{3}($P$_{4})_{20}$} & $5'825'561$ & $34$ & $9.7\times 10^{-3}$ - $1.0\%$ & $6.2$ & $26.0$ & $1$-$14$  & $1.7$ & $4.8\times 10^{-4}$\\

                    &  & $6'503'837$ & $288$ & $7.4\times 10^{-2}$ - $8.4\%$ & $11.5$ & $54.9$ & $1$-$85$ & $1.0$ & $2.6\times 10^{-4}$\\
\hline
\multirow{2}{*}{L3} & \multirow{2}{*}{$($C$_{1})_{4}($P$_{4})_{30}$} & $5'829'183$ & $24$ & $6.8\times 10^{-3}$ - $1.0\%$ & $7.4$ & $6.9$ & $2$-$11$  & $0.0$ & $-$\\

                    &  & $6'386'342$ & $194$ & $5.0\times 10^{-2}$ - $8.3\%$ & $15.3$ & $133.7$ & $2$-$70$ & $0.0$ & $-$\\
\hline
T1 & $($C$_{1})_{3}($P$_{4})_{10}$ & $6'506'929$ & $520$ & $1.3\times 10^{-1}$ - $8.9\%$  & $11.1$ & $54.8$ & $1$-$50$ & $3.9$ & $1.0\times 10^{-3}$\\
\hline
T2 & $($C$_{1})_{3}($P$_{4})_{20}$ & $6'487'350$ & $288$ & $7.4\times 10^{-2}$ - $8.4\%$  & $9.3$ & $41.7$ & $1$-$39$ & $4.2$ & $1.1\times 10^{-3}$\\                  
\hline\hline
\end{tabular}
\caption{\label{tab:stat_bip}
Basic statistics regarding micelle size for the equilibration of CG water with surfactants at different concentrations.
In all cases, the standard deviation of the volume is always around $1\%$ of the average value. $C$ stands for
concentration.}
\end{table*}
%=======================================
For the surfactant L2, L3, T1 and T2, we apply the same simulation settings with the 
exception that the timestep size is reduced to $10$ [fs] for stability reasons 
related to bond interactions of the surfactant molecules. The temperature is always kept 
fixed at $T=298$ [K] and the initial state consists of a given number of surfactant 
molecules, for the concentrations of Tab.~\ref{tab:stat_bip}, between two slabs of water beads 
($50'000$ in total). For the typical experimental concentration of $1$ wt$\%$, the RDFs of Fig.~\ref{fig:rdf_bip}
indicate that the tail beads, for the surfactant L2, are separated by larger distances because
the first maximum is lower. From the size distributions of micelles 
plotted in Fig.~\ref{fig:freq_mic}, it can be seen that the linear surfactants tend to form larger micelles on
average. A summary of micelle statistics is given in Tab.~\ref{tab:stat_bip};
in all cases we used $r_{\mathrm{min}}=7$ [\AA] (cf.~Fig.~\ref{fig:rdf_bip}).
%=======================================
%=======================================

%=======================================
\begin{table*}[t]
\begin{tabular}{l|l|c|c|c|c|c|c|c|c}
\hline\hline
Surfactant & Mapping & no.~water beads & no.~surfactants & $C$ [wt$\%$] & $N_{\mathrm{micelle}}$ & Size range & no.~monomers& $d$ [molecules/\AA$^{3}$] & $\theta$ [$^{\circ}$]\\
\hline
\multirow{2}{*}{L2} & \multirow{2}{*}{$($C$_{1})_{3}($P$_{4})_{20}$}  & $68'727$ & $43$  & $0.9\%$  & $17.0$  & $1$-$14$ & $12.1$ & $0.0074$ & $131.1^{\circ}$\\

                    &  & $60'351$ & $359$ & $8.7\%$ & $108.3$ & $1$-$52$ & $55.5$ & $0.0065$ & $130.7^{\circ}$\\
\hline
\multirow{2}{*}{L3} & \multirow{2}{*}{$($C$_{1})_{4}($P$_{4})_{30}$} & $68'690$ & $28$  & $0.9\%$  & $10.3$  & $1$-$10$ & $6.6$  & $0.0074$ & $131.1^{\circ}$\\
 
                    &  & $63'555$ & $239$ & $8.1\%$ & $36.6$ & $1$-$66$ & $11.3$ & $0.0068$ & $131.2^{\circ}$\\
\hline\hline
\end{tabular}
\caption{\label{tab:cylinder}
Characteristics of the cylindrical droplets of initial radius $130$ [\AA] containing linear surfactants. $C$ indicates the concentration.
The average number of micelles is designated by $N_{\mathrm{micelle}}$.
The cutoff distance for the extraction of micelle clusters is $7$ [\AA] (see Fig.~\ref{fig:rdf_bip}).}
\end{table*}
%=======================================

%=======================================
\begin{table*}[t]
\begin{tabular}{l|l|c|c|c|c|c|c|c}
\hline\hline
Surfactant & Mapping & no.~surfactants & $C$ [wt$\%$] & $N_{\mathrm{micelle}}$ & Size range & no.~monomers & $d$ [molecules/\AA$^{3}$] & $\theta$ [$^{\circ}$]\\
\hline
\multirow{3}{*}{L1} & \multirow{3}{*}{$($C$_{1})_{3}($P$_{4})_{10}$}  & $85$  & $1.1\%$  & $41.8$ & $1$-$12$ & $21.6$ & $0.0076$ & $129.6^{\circ}$\\

                    &  & $424$ & $5.2\%$  & $112.4$ & $1$-$37$ & $46.3$ & $0.0076$ & $118.2^{\circ}$\\

                    &  & $847$ & $10.5\%$ & $112.3$ & $1$-$165$ & $40.6$ & $0.0076$ & $114.5^{\circ}$\\
\hline
\multirow{3}{*}{L2} & \multirow{3}{*}{$($C$_{1})_{3}($P$_{4})_{20}$}  & $39$  & $0.8\%$  & $27.4$ & $1$-$9$  & $19.4$ & $0.0076$ & $130.1^{\circ}$\\

                    &  & $196$ & $4.2\%$  & $75.1$ & $1$-$21$ & $35.4$ & $0.0075$ & $123.5^{\circ}$\\

                    &  & $393$ & $8.4\%$ & $91.1$ & $1$-$40$ & $33.9$ & $0.0074$ & $111.7^{\circ}$\\
\hline
\multirow{4}{*}{L3} & \multirow{4}{*}{$($C$_{1})_{4}($P$_{4})_{30}$}  & $26$  & $0.8\%$  & $15.1$ & $1$-$7$  & $8.5$  & $0.0076$ & $130.5^{\circ}$\\
 
                    &  & $132$ & $4.1\%$  & $32.1$ & $1$-$21$ & $6.7$  & $0.0075$ & $125.0^{\circ}$\\
 
                    &  & $265$ & $8.3\%$ & $34.3$ & $1$-$43$ & $3.7$  & $0.0072$ & $114.9^{\circ}$\\

                    &  & $530$ & $16.7\%$ & $65.0$ & $1$-$206$ & $28.9$ & $0.0065$ & $124.7^{\circ}$\\
\hline
\multirow{2}{*}{T1} & \multirow{2}{*}{$($C$_{1})_{3}($P$_{4})_{10}$}  & $85$  & $1.1\%$ & $44.4$ & $1$-$13$ & $24.1$ & $0.0076$ & $129.2^{\circ}$\\

                    &  & $424$ & $5.2\%$  & $119.1$ & $1$-$33$ & $45.4$ & $0.0076$ & $116.0^{\circ}$\\
\hline
\multirow{3}{*}{T2} & \multirow{3}{*}{$($C$_{1})_{3}($P$_{4})_{20}$} & $39$  & $0.8\%$  & $28.3$ & $1$-$7$ & $21.0$ & $0.0076$ & $129.8^{\circ}$\\

                    &  & $196$ & $4.2\%$  & $66.3$ & $1$-$20$ & $23.4$ & $0.0075$ & $122.1^{\circ}$\\

                    &  & $393$ & $8.4\%$ & $88.3$ & $1$-$36$ & $22.2$ & $0.0074$ & $108.3^{\circ}$\\
\hline
                T3  & $($C$_{1})_{5}($P$_{4})_{20}$ & $212$  & $5.0\%$ & $43.8$ & $1$-$30$ & $9.6$ & $0.0075$ & $125.1^{\circ}$\\
\hline\hline
\end{tabular}
\caption{
Results for the cylindrical droplets of initial radius $130$ [\AA] with the surfactants arranged in 
the initial configuration in the proximity of contact line; all droplets contain $68'255$ water beads. 
$C$ stands for concentration; $N_{\mathrm{micelle}}$ is the average number of micelles, and size range
is the aggregation number variation. Micelle clusters are extracted using a cutoff distance of $7$ [\AA] 
for all surfactants (see Fig.~\ref{fig:rdf_bip}).
The simulations for the systems containing linear surfactants with a concentration higher than $1$ wt$\%$ lasted for 100 [ns] 
and the analysis is performed on data collected under the same conditions over the last $40$ [ns]. This choice is dictated by 
the fact that longer evolutions are necessary in order to reach a state near the equilibrium in these cases. The contact angle 
for the surfactant L3 at $16.7$ wt$\%$ is not in line with the results for the lower concentrations suggesting that the system is 
not yet well equilibrated. For these longer dynamics, the stratification of water did not occur. 
\label{tab:contact_line}}
\end{table*}
%=======================================
%=======================================
\begin{figure*}[t]
\includegraphics[width=5cm]{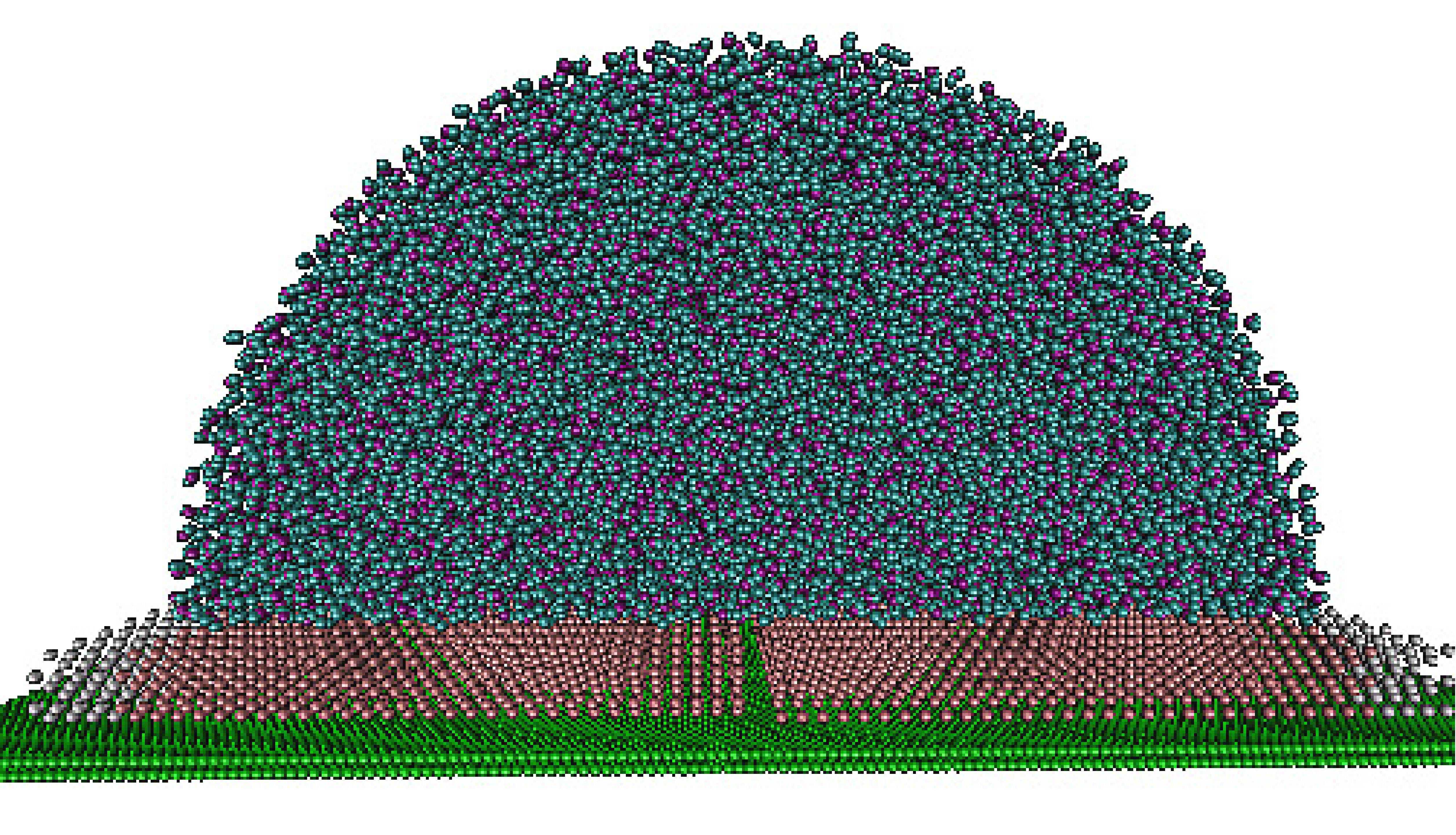}\hspace{0.5cm}
\includegraphics[width=5cm]{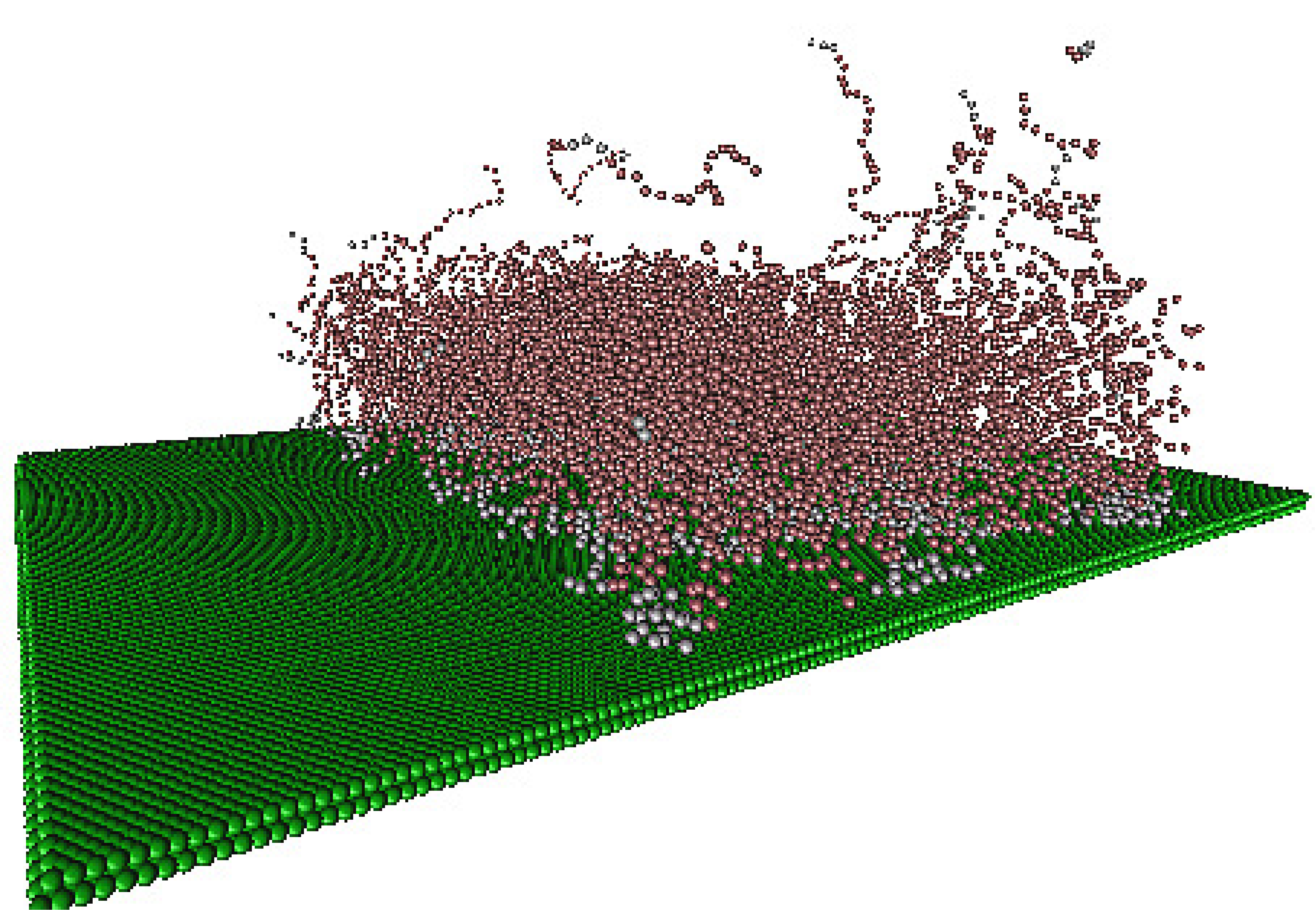}\hspace{0.5cm}
\includegraphics[width=5cm]{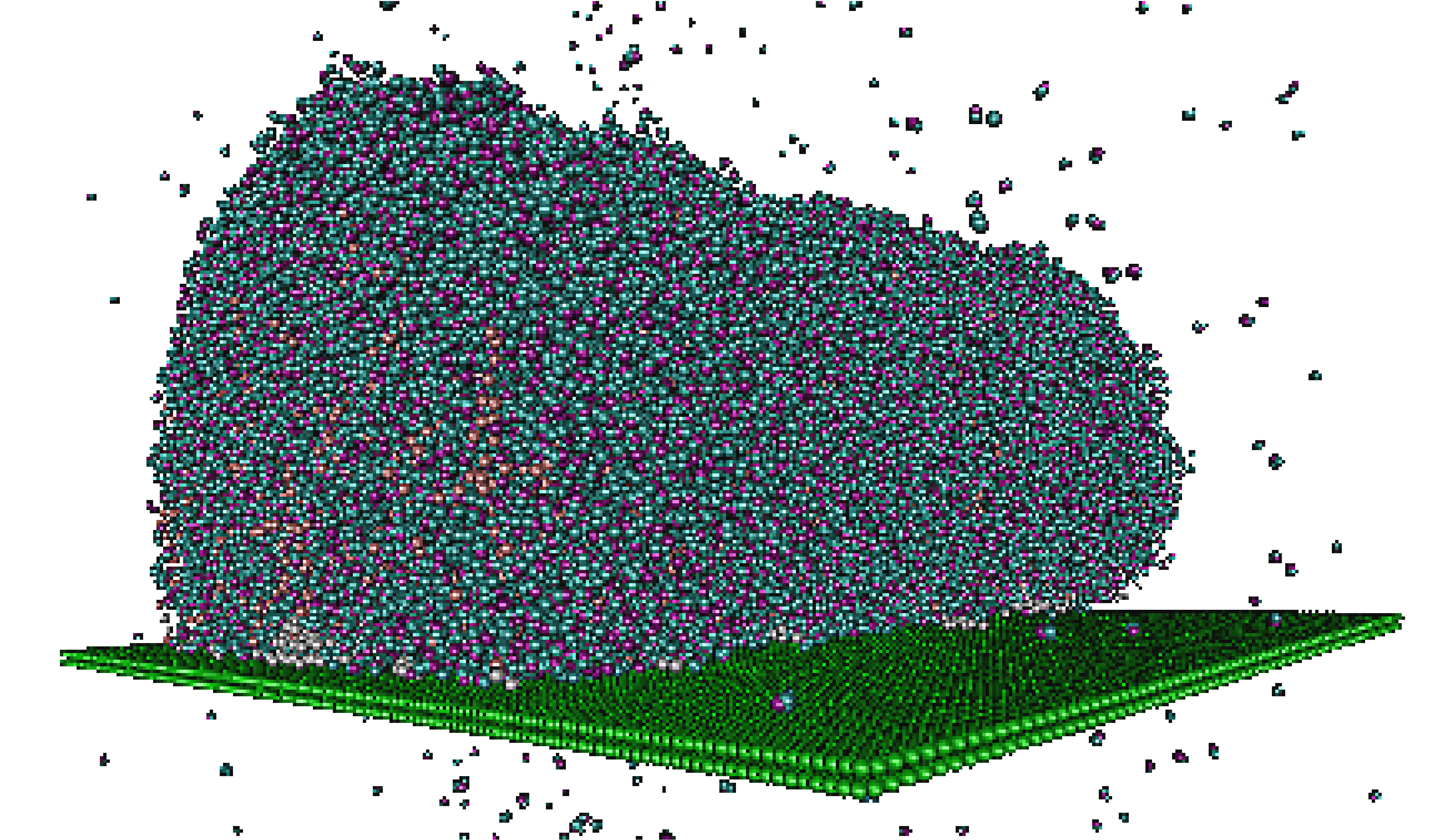}
\caption{
(Color online) From Left to Right: Droplet with surfactant L3 at $8$ wt$\%$ (see Tab.~\ref{tab:contact_line}) 
in the initial configuration, final configuration without water beads and final configuration including
water beads. At equilibrium, the surfactant are oriented with their head toward the fluid.
\label{fig:drp_surf}}
\end{figure*}
%=======================================
%=======================================
\begin{figure*}[t]
\includegraphics[width=8.5cm]{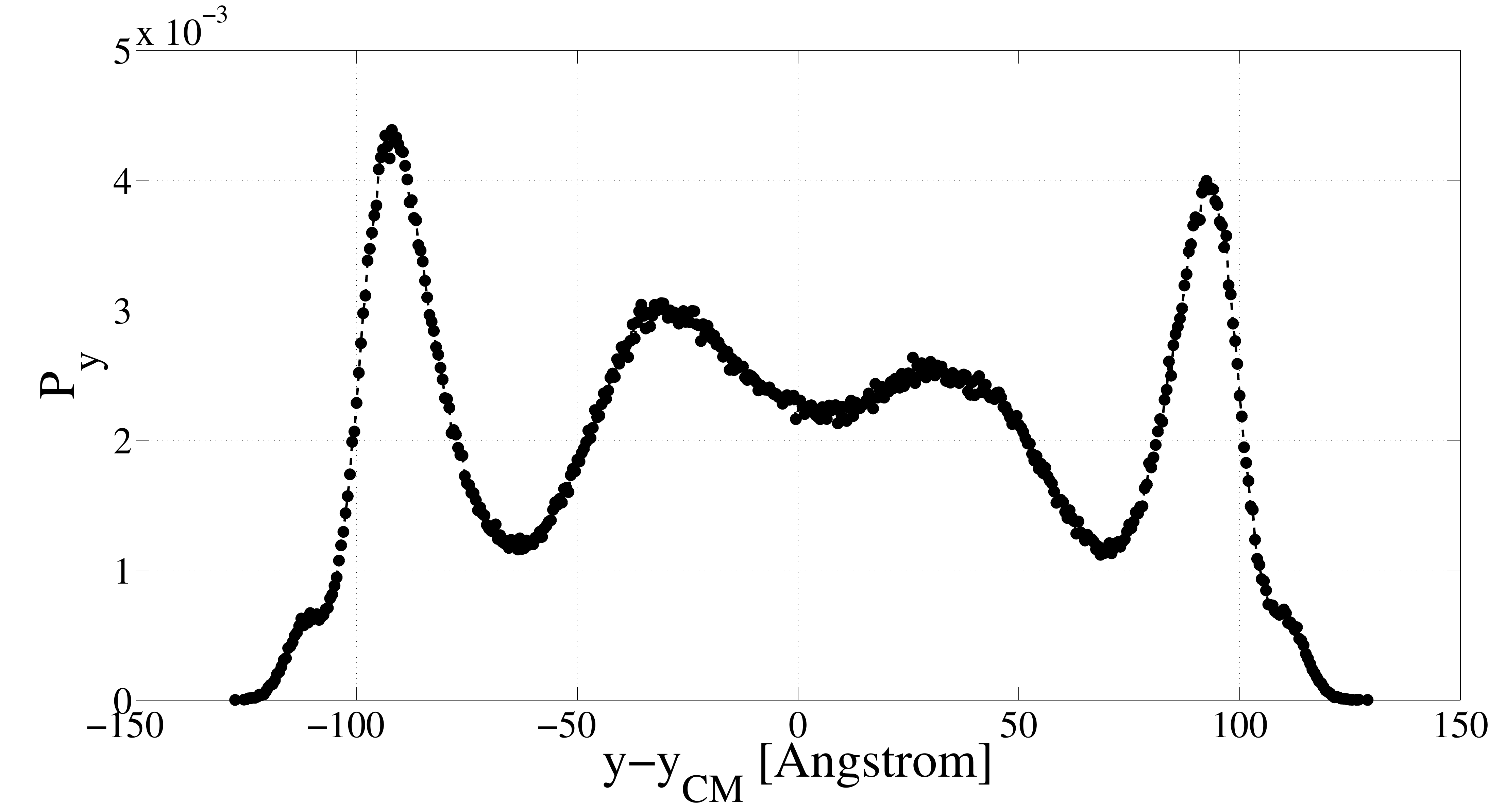}\hspace{0.5cm}
\includegraphics[width=8.5cm]{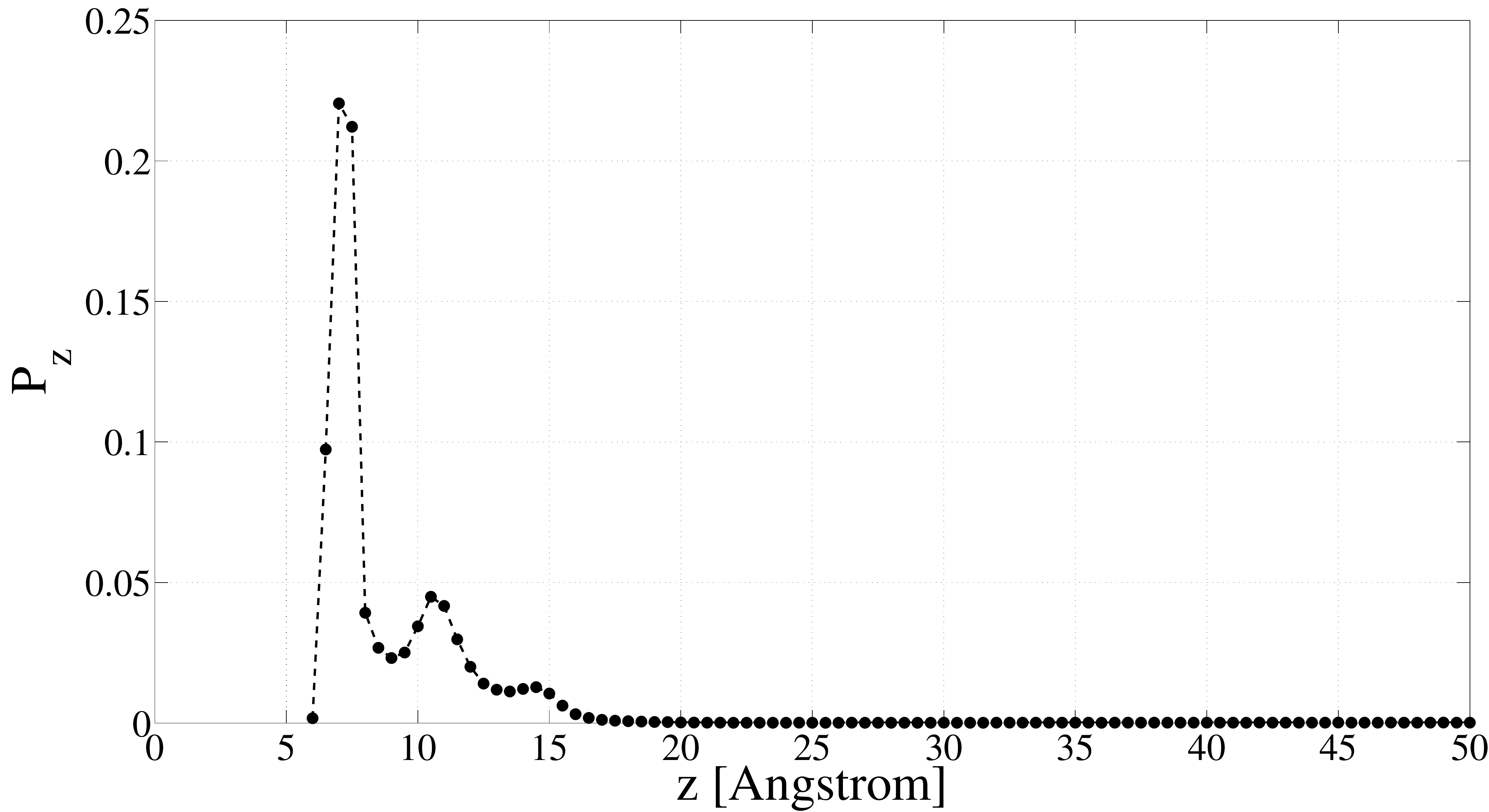}\\
\includegraphics[width=8.5cm]{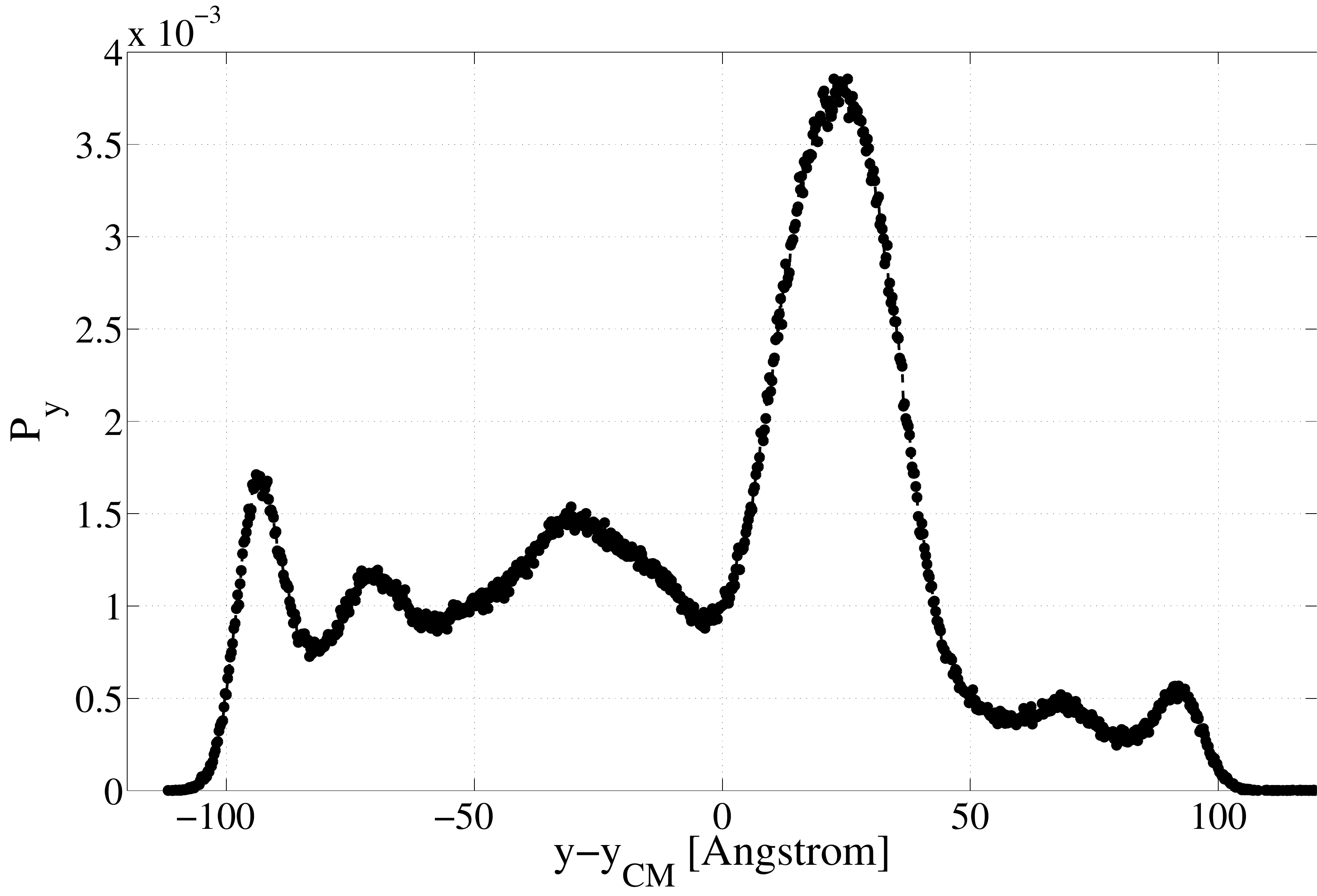}\hspace{0.5cm}
\includegraphics[width=8.5cm]{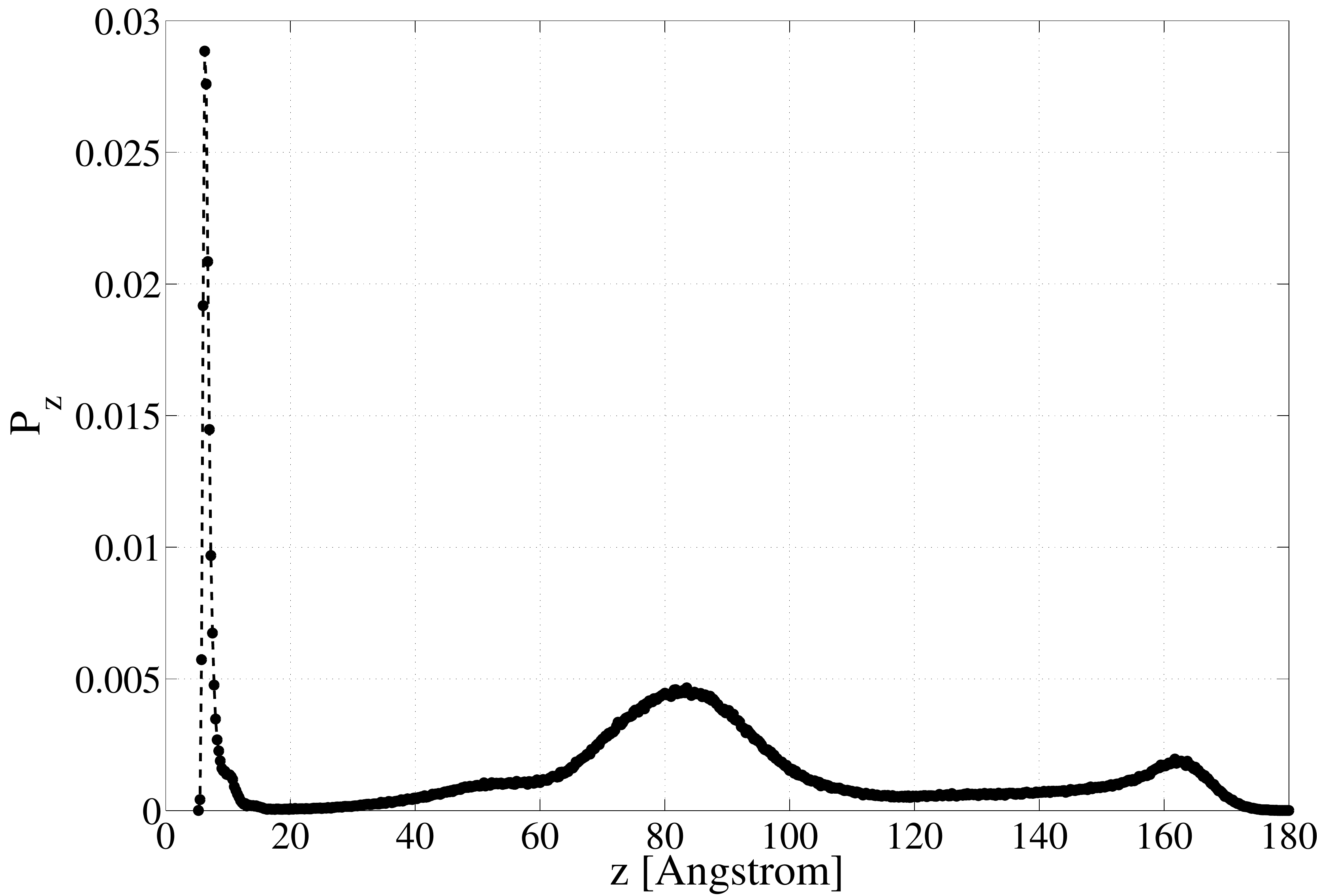}
\caption{
Distributions of C$_{1}$ beads in space for the cylindrical droplets containing surfactants L2
at concentration $8$ wt$\%$.
Top: Results for the droplet prepared as shown in Fig.~\ref{fig:drp_surf}, with the surfactants in the
proximity of contact line. Many surfactants are near the contact line, i.e.~$\mid y-y_{\mathrm{CM}}\mid\approx 90$ 
[\AA], and at the solid-liquid interface. Bottom: Results obtained from an equilibrated solution 
(cf.~Tab.~\ref{tab:cylinder}). The position and height of the peaks indicate that a significant number 
of surfactants is in the bulk of the droplet. Similar conclusions can be inferred from the outcomes 
for the other case studies. \label{fig:droplet_surf}}
\end{figure*}
%=======================================
\begin{figure*}[t]
\includegraphics[width=8.5cm]{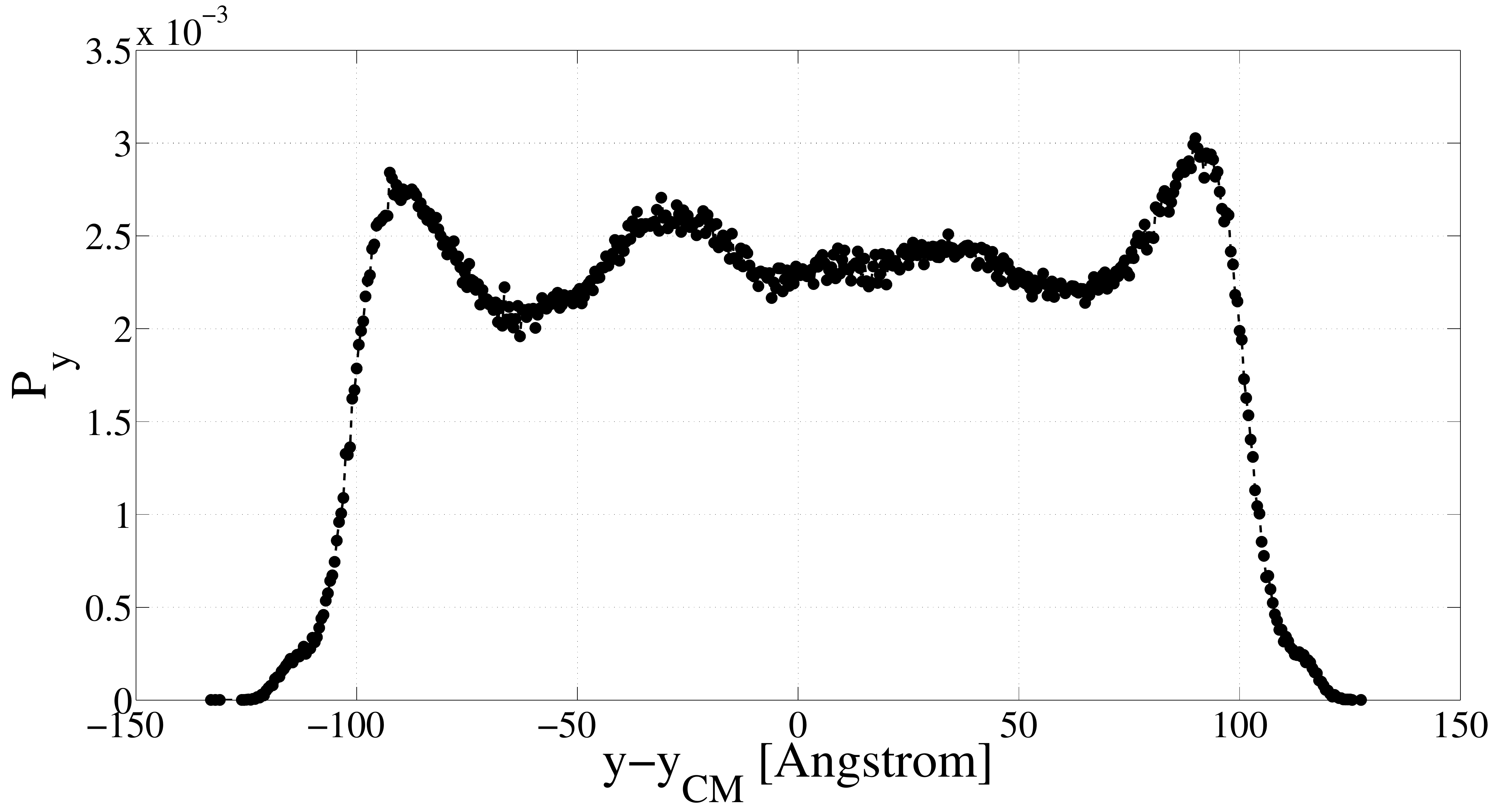}\hspace{0.5cm}
\includegraphics[width=8.5cm]{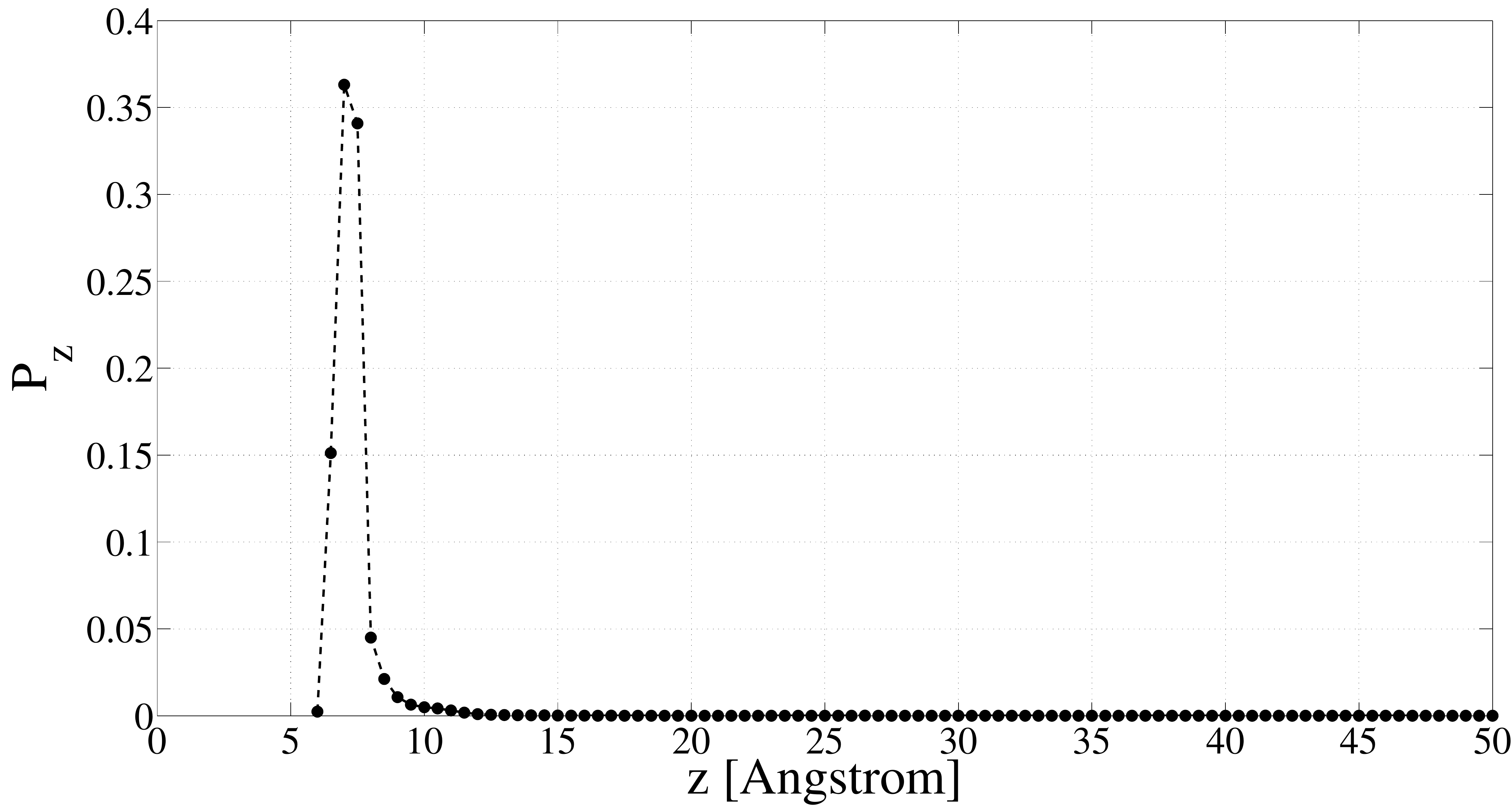}
\caption{
Spatial distributions of C$_{1}$ tail beads for the cylindrical droplet containing surfactant T2
at $8$ wt$\%$ (see Tab.~\ref{tab:contact_line}). In this case, the surfactants are no longer neatly 
localized along the contact line. It also appears that the adsorption layer is more uniform 
(cf.~Fig.~\ref{fig:droplet_surf}). The plots for the other T-shaped surfactants lead to the same
conclusions. \label{fig:Tshaped}}
\end{figure*}
%=======================================
\begin{figure*}[t]
\includegraphics[width=8.5cm]{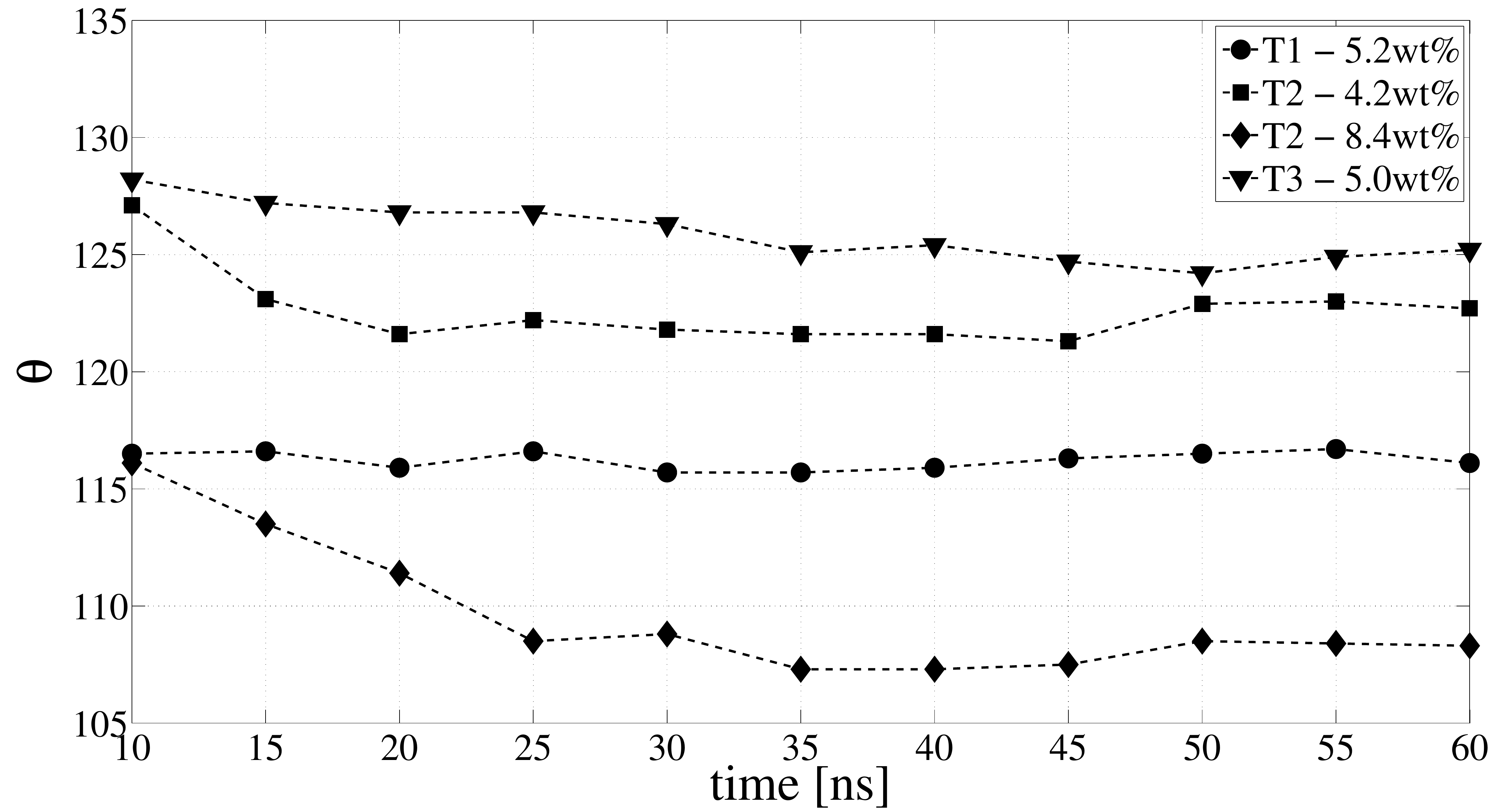}\hspace{0.5cm}
\includegraphics[width=8.5cm]{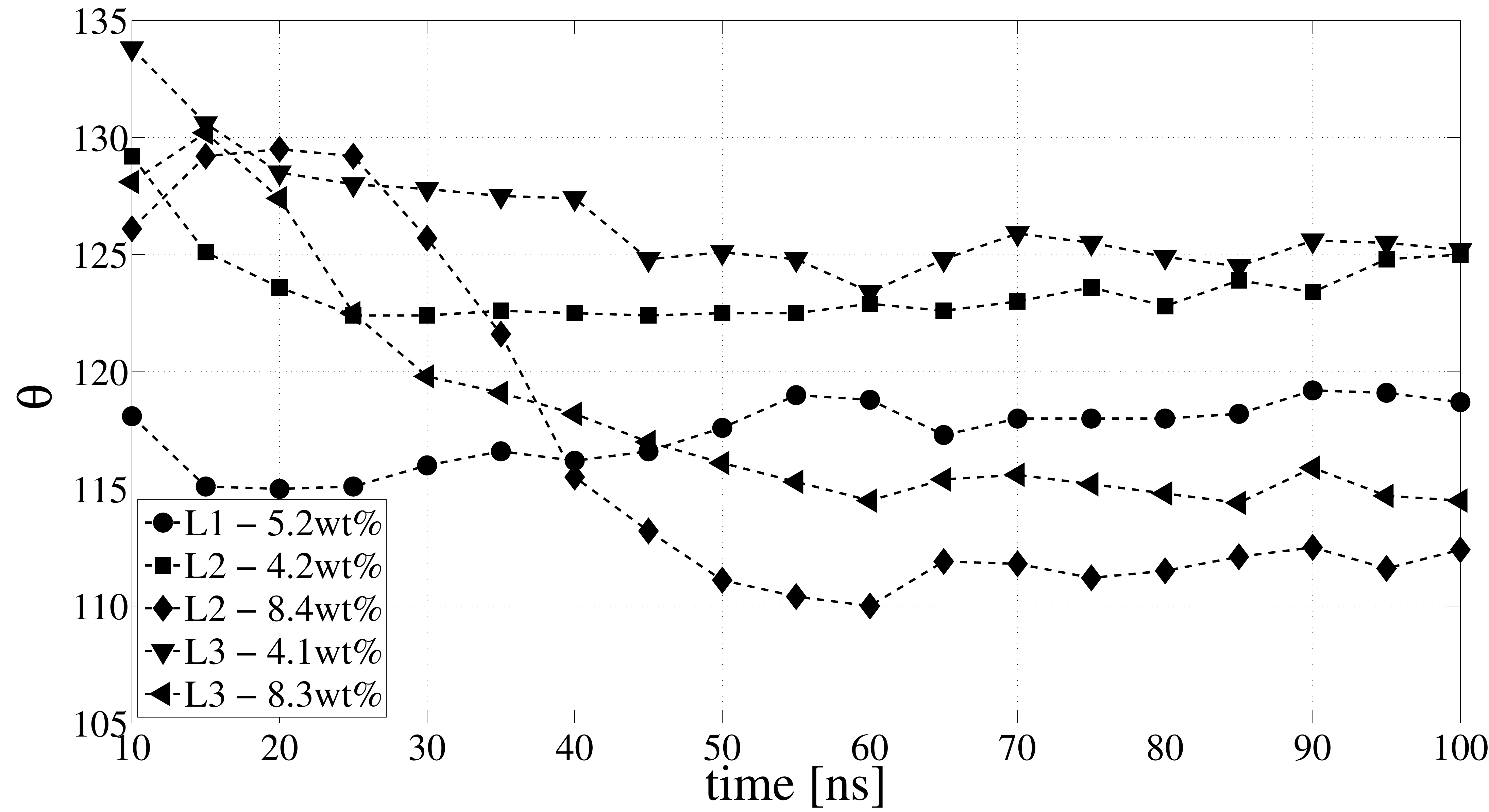}
\caption{
Contact angle $\theta$ in the course of time for the cylindrical droplets of Tab.~\ref{tab:contact_line}.
In general, T-shaped surfactants spread faster. The analysis uses the data collected in time intervals
of $5$ [ns]. \label{fig:spreading}}
\end{figure*}
%=======================================

\textit{Wetting with surfactants}.~Also for this train of simulations, the timestep size
is reduced to $10$ [fs]. In all other respects, the settings used for cylindrical droplets
of pure water are maintained. The effect of the presence of
surfactants is to lower the contact angle and this is possible if the tail beads, 
displaying hydrophobic behavior, interact more strongly with the CG graphene than the head beads P$_{4}$.
For this reason, it is assumed that the interaction strength is $2\times \varepsilon_{\mathrm{CW}}=0.115$ [kcal/mol].
This approach is chemically consistent since, in graphene, carbon atoms have no polarity.
The parameter $\sigma$ for the LJ interactions with graphene beads is again $6.24$ [\AA].
The results of Tab.~\ref{tab:cylinder} indicate that the contact angle is almost insensitive to 
significant variations of surfactant concentration using equilibrated solutions. 
Since in general it is observed a stronger micellization process,
we repeat the previous simulations with cylindrical droplets of pure water and the surfactants arranged
between water and graphene as illustrated in Fig.~\ref{fig:drp_surf}. In this way, the solution is not yet
properly equilibrated, but more surfactant molecules are closer to the expected configuration, 
that is, near the contact line \cite{york,york2,shen}. At $1$ wt$\%$, all the systems still do not experience
any influence of the presence of surfactants (see Tab.~\ref{tab:contact_line}). 
It is worth noting that, in similar studies \cite{baoukina,zwitterionic}, low surfactant
concentrations resulted in a transient increase of the surface tension for
the MARTINI model.
On the other hand, the concentration
around $4$ wt$\%$ provides a basis of comparison among the different systems. The results for linear surfactants 
prove that, in first approximation, the dominant factors promoting wetting are the length and apolarity of the
hydrophobic tail. Indeed, comparison between surfactants L2 and L3 indicate that one single C$_{1}$ bead
is sufficient to counteract the effect of ten P$_{4}$ beads (their contact angles differ by a few degrees).
We want to highlight the preferred arrangement of surfactant molecules inside the droplets. To this end, 
it is useful to look at the spatial distributions $P_{y}(y-y_{\mathrm{CM}})$ and $P_{z}(z)$ for C$_{1}$ beads 
along the $y$ and $z$ axes, as the notation suggests. Here $y_{\mathrm{CM}}$ is the coordinate of the center of 
mass of CG water. As evidenced by Fig.~\ref{fig:droplet_surf}, linear surfactant molecules are principally
located at the solid-liquid interface. Their organization is not uniform, as indicated by the
statistics for micelles of Tab.~\ref{tab:contact_line}. The weak tendency to form layers, together with 
the fact that the distribution $P_{y}$ is strongly peaked around the contact line, give further support to the hypothesis 
that the number of hydrophobic units is the dominant factor for enhanced wetting. Regarding the T-shaped topology,
from Fig.~\ref{fig:Tshaped} it is seen that less surfactant molecules accumulate along the contact line. The fact
that the contact angle remains almost unchanged, as compared to the linear counterparts (see Tab.~\ref{tab:contact_line}), 
let us conclude that their action is more effective. In macroscopic systems, 
they should thus wet better than linear surfactants, since the plateau in Fig.~\ref{fig:Tshaped} is already almost
at the same height of the peaks near the contact line. Interestingly, a longer hydrophobic tail does not lead
to a lower contact angle. It thus arises that for T-shaped surfactants a shorter hydrophilic headgroup can
enhance wetting more than a longer tail. The plots of Fig.~\ref{fig:spreading} shows that one
advantage of the T-shaped topology is faster spreading. After $25$ [ns] of dynamics, the contact angles for T-shaped
surfactants vary at most of $1.7^{\circ}$ (surfactant T2 at $4.2$ wt$\%$). In the same evolution period,
with the exception of the surfactant L2 at $4.2$ wt$\%$, the variations of the contact angle for linear surfactants
are more marked: at least of $3.9^{\circ}$ (surfactant L1 at $5.2$ wt$\%$) and at most of $19.2^{\circ}$ (surfactant L2 at $8.4$ wt$\%$).
The influence of the topology for enhanced spreading was already recognized in the literature \cite{york,york2,shen}.
%======================================================================

\section{Conclusions}

Small concentrations of surfactants around $1$ wt$\%$ can cause significant reductions of the 
contact angle. As an example, for the surfactant Triton\textsuperscript{\textregistered} X-100,
the surface tension of water of $72.5$ [mN/m] decreases linearly with concentration
up to $0.03$ wt$\%$, when the  minimum of $31$ [mN/m] is attained; for higher concentrations
the surface tension remains of course constant. It turns out that the proper treatment of wetting
phenomena by aqueous surfactant solutions requires conditions of difficult realization by the
present computational capabilities. Coarse-grained models allowed us to move one step further
toward a more realistic representation of such systems in terms of surfactant concentrations.
At the typical experimental concentration of $1$ wt$\%$, the coarse-grained MARTINI force field
\cite{martini1,martini2} marks no difference between the surfactants investigated here.  
For higher concentrations there appears that it is possible to discriminate the wetting behavior 
of the various surfactants. The main instrumental conclusion of our study is that the length and 
apolarity of the hydrophobic tail determine to a larger extent the wetting behavior for the linear topology. 
Instead, the length of the hydrophilic headgroup appears to be more relevant for the T-shaped topology. 
In the framework of our simulations, the T-shaped 
topology does not lead to a substantial decrease of contact angle vis-a-vis linear surfactants. Lennard-Jones forces 
are short range and the reduction of the contact angle is essentially driven by the accumulation of surfactants
along the contact line. Micelle statistics demonstrates that linear surfactants pack more
tightly, compensating the fact that the hydrophobic tail beads of T-shaped surfactants are
on average closer to the graphitic substrate. It thus follows that the stronger self-assembly behavior
of surfactants can also favor wetting. On the other hand, the weaker micellization of T-shaped
surfactants can result in faster spreading. Finally, our parameterization of surfactants
while very simple can encompass and provide benchmarks for several commercially available
surfactants. More comparative work remains to be done and our results offer some insight
on the complex interplay between micellization, spreading and wetting.
%===========================================================================

\begin{acknowledgments}
This work was supported by the Swiss Innovation Promotion Agency (KTI/CTI)
through BiPCaNP project under grant P.~No.~10055.1. Computations were done with the
facilities of CSCS and iCIMSI-SUPSI. We are grateful to their staff for assistance.
\end{acknowledgments}

%===========================================================================

%=======================================================================
\end{document}